\documentclass[12pt]{iopart}

\usepackage[sort&compress,numbers]{natbib}
\usepackage{url}
\usepackage{iopams}  

\expandafter\let\csname equation*\endcsname\relax

\expandafter\let\csname endequation*\endcsname\relax

\usepackage{amsmath}
\usepackage{graphics}
\usepackage{graphicx}
\usepackage{placeins}
\usepackage{esint}
\usepackage{hyperref}
\usepackage{comment}
\setcitestyle{square}

\makeatletter
\newsavebox{\@brx}
\newcommand{\llangle}[1][]{\savebox{\@brx}{\(\m@th{#1\langle}\)}%
  \mathopen{\copy\@brx\kern-0.5\wd\@brx\usebox{\@brx}}}
\newcommand{\rrangle}[1][]{\savebox{\@brx}{\(\m@th{#1\rangle}\)}%
  \mathclose{\copy\@brx\kern-0.5\wd\@brx\usebox{\@brx}}}
\makeatother

\begin{document}

\title[Thermodynamic geometry of ideal quantum gases]{Thermodynamic geometry of quantum gases: \\
a general framework and a geometric picture of BEC-enhanced heat engines}
\author{Joshua Eglinton$^{1,2}$, Tuomas Pyhäranta$^{3}$, Keiji Saito$^{4}$ and Kay Brandner$^{1,2}$}

\address{
$^{1}$School of Physics and Astronomy,
University of Nottingham,
Nottingham NG7 2RD,
United Kingdom\\
$^{2}$Centre for the Mathematics and Theoretical Physics of Quantum Non-equilibrium Systems,
University of Nottingham,
Nottingham NG7 2RD, 
United Kingdom\\
$^{3}$QTF Center of Excellence, Department of Applied Physics, Aalto University, P. O. Box 11000, FI-00076 Aalto, Espoo, Finland\\
$^{4}$Department of Physics, Keio University, 3-14-1 Hiyoshi, Kohoku-ku, Yokohama 223-8522, Japan}
\ead{joshua.eglinton@nottingham.ac.uk}
\vspace{10pt}
\begin{indented}
\item[]December 2022
\end{indented}

\begin{abstract}
Thermodynamic geometry provides a physically transparent framework to describe thermodynamic processes in meso- and micro-scale systems that are driven by slow variations of external control parameters. Focusing on periodic driving for thermal machines, we extend this framework to ideal quantum gases. To this end, we show that the standard approach of equilibrium physics, where a grand-canonical ensemble is used to model a canonical one by fixing the mean particle number through the chemical potential, can be extended to the slow driving regime in a thermodynamically consistent way. As a key application of our theory, we use a Lindblad-type quantum master equation to work out a dynamical model of a quantum many-body engine using a harmonically trapped Bose-gas. Our results provide a geometric picture of the BEC-induced power enhancement that was previously predicted for this type of engine on the basis of an endoreversible model [New J. Phys. $\textbf{24}$, 025001 (2022)]. Using an earlier derived universal trade-off relation between power and efficiency as a benchmark, we further show that the Bose-gas engine can deliver significantly more power at given efficiency than an equally large collection of single-body engines. Our work paves the way for a more general thermodynamic framework that makes it possible to systematically assess the impact of quantum many-body effects on the performance of thermal machines.  
\end{abstract}

%
%
%
%
%

\section{Introduction}
Thermodynamic processes on meso- and microscopic length and energy scales are commonly described in terms of non-autonomous differential equations, that couple the state variables of a working system such as the elements of a density matrix with external control variables like magnetic fields \cite{Seifert2012,Vinjanampathy2016}. Thermodynamic quantities such as net work output or net entropy production depend on both state and control variables. Thus, any attempt to optimize a given process with respect to its driving protocols generically leads to a complicated dynamical control problem. For systems that are weakly coupled to a thermal environment and driven slowly, relative to their internal relaxation timescale, thermodynamic geometry provides an elegant method to simplify such problems \cite{Sivak2012,Crooks2012,Rotskoff2015,Zulkowski2015,Zulkowski2015b,Cavina2017,Miller2019,Abiuso2020,Blaber2020,blaber2022optimal,Mehboudi2022,Abiuso2022,Scandi2022}. The key idea of this approach is to solve the equations of motion of the working system, by means of adiabatic perturbation theory \cite{Cavina2017,Weinberg2017}. The state variables of the system thereby become functions of the control parameters and their time derivatives. Upon inserting these solutions into the expressions for the considered figures of merit, they become functionals of the control variables and their time derivatives, which can be optimized by solving the corresponding Euler-Lagrange equations without dynamical constraints. In this way, thermodynamic quantities can be related to geometric concepts. For instance, the dissipation of a given process, when minimized with respect to the parameterization of the applied control protocols, can be expressed as the length of the control path with respect to a system-dependent pseudo-Riemannian metric in the space of control parameters; this quantity, which was first introduced in macroscopic thermodynamics \cite{Weinhold1975,Andresen1984,Brody1995,Ruppeiner1995,Salmon1983}, is known as thermodynamic length \cite{Crooks2007,Machta2015,Scandi2019}. 

Its universal structure, which does not depend on the specific equations of motion describing the working system, and its ability to consistently account for finite-time corrections, make thermodynamic geometry a powerful tool to analyze small-scale thermal machines operating close to equilibrium \cite{Brandner2020,Frim2022,Miller2020,Erdman2022,Lu2022,Bhandari2020,Potanina2021,Eglinton2022,Izumida2021,Izumida2022,Alonso2022,Raz2016,Bhandari2022,Frim2022b,Hino2021,Hayakawa2021,Chen2022}. 
Recent results derived in this framework include universal bounds on the efficiency, power and constancy \cite{Pietzonka2018}, i.e., inverse power fluctuations, of small-scale heat engines driven by a periodically varying heat source \cite{Brandner2020,Frim2022,Miller2020} or a fixed temperature gradient between two thermal reservoirs \cite{Lu2022,Bhandari2020,Potanina2021,Eglinton2022,Alonso2022,Izumida2021,Izumida2022}. In addition, a whole variety of methods have been developed to find optimal control protocols for small-scale thermal machines which, for example, maximize power output and efficiency \cite{Raz2016,Alonso2022,Bhandari2022,Frim2022b} or suppress power fluctuations \cite{Miller2020,Lu2022,Erdman2022}. 

A promising avenue towards scaling up the power and constancy of quantum thermal machines is to replace working systems with few degrees of freedom by many-body systems capable of hosting collective effects, which may lead to uncovering new mechanisms of energy conversion \cite{Chen2017,Li2018,Boubakour2022,Niedenzu2018,Kloc2019,Yadin2022,Kolisnyk2022,Carollo2020,jaseem2022,Ma2017,Fogarty_2021,wen2022,Revathy2022,Myres2020,Myres2021,Myres2022,Koch2022,marzolino2022,Skelt2019,Carollo2020b,Mukherjee2021,Mayo2022,Li2022,solfanelli2022}. Such effects, whose thermodynamics is yet to be fully understood, include: tunable interactions between particles, which can be used for work-extraction \cite{Chen2017,Li2018,Boubakour2022}; super-radiance and broken time-translation symmetry, which emerge in multi-level systems coupled to a thermal bath via collective observables \cite{Niedenzu2018,Kloc2019,Yadin2022,Kolisnyk2022,Carollo2020,jaseem2022}; quantum phase transitions \cite{Ma2017,Fogarty_2021,wen2022,Revathy2022} or quantum statistics, which provides a means of controlling an effective pressure that has no classical counterpart \cite{Myres2020,Myres2021,Myres2022,Koch2022,marzolino2022}. Thermodynamic geometry offers a powerful tool to analyze these phenomena from a unifying perspective and thus a potential avenue towards a universal framework describing how many-body effects can alter the performance of quantum thermal machines.

Here, we take a first step in this direction by investigating the formation of Bose-Einstein condensates (BECs) in ideal quantum gases as an example of a quantum many-body effect which has been recently shown to enhance the performance of heat engines under suitable conditions \cite{Myres2022}. There, an endoreversible Otto cycle is considered for a harmonically confined Bose gas as a working medium. Cycles crossing the transition, between the condensate and normal phase are shown to enhance engine performance. This effect arises from the reduced effective pressure of the working system in the condensate phase, which makes it possible to carry out a compression stroke with significantly lower work input than in the normal phase. Expanding the gas in the normal phase, thus leads to an enhanced net work output compared to an engine that operates solely in the condensed or normal phase. To analyze this effect through the lens of thermodynamic geometry, we develop a dynamical framework to describe arbitrary thermodynamic cycles close to equilibrium with ideal quantum gases. The central idea underpinning this framework is that the principle of equivalence of ensembles should still hold in the adiabatic-response regime, where the system is not driven far away from equilibrium. 

Specifically, we start with a grand-canonical picture, where the chemical potential and temperature of a thermal reservoir coupled to the working system are considered as external control parameters along with the mechanical parameters that enter its Hamiltonian. In this setting, we fix the chemical potential such that at each instance of time the average number of particles within the system remains constant, thus making the chemical potential dependent on the remaining control variables. This procedure, which we show to be thermodynamically consistent, allows us to construct an effectively canonical picture. It thus becomes possible to develop the thermodynamic geometry of quantum-gas engines in analogy with earlier approaches based on few-body systems. As a result, the thermodynamic impact of Bose-Einstein condensation as a quantum many-body effect can be quantitatively assessed. We illustrate this idea for a Stirling cycle, whose working fluid consists of an ideal Bose gas in a tunable harmonic trap. For this model, we analyse the BEC-induced work enhancement as well as the trade-off between power and efficiency \cite{Brandner2020}. Our results show that although both the work output and minimum dissipation are increased by the BEC, for large system sizes, the power output at given efficiency can be increased compared to a collection of single-body engines.

We organise this work in the following sections. In Sec.~\ref{sec: Framework} we develop our geometric framework for the description of ideal quantum gases in the adiabatic-response regime. We continue by showing that the adiabatic-response coefficients in the grand-canonical setting gives rise to thermodynamically consistent, effective adiabatic-response coefficients in the canonical picture once the mean particle number of the system is fixed via the chemical potential. These effective coefficients enter thermodynamic vector potentials, Berry curvatures and a thermodynamic metric, which depends only on the remaining control variables, i.e., the temperature of the reservoir and the mechanical parameters controlling the Hamiltonian of the system. In Sec.~\ref{sec: MBHE}, we apply our framework to a many-body heat engine, whose working system consists of an ideal Bose gas undergoing a Stirling cycle and analyze the thermodynamic effect of Bose-Einstein condensation. Finally, we provide our concluding remarks and outlook in Sec.~\ref{sec: Discussion}.

\section{Framework}
\newcommand{\Lmd}{\boldsymbol{\Lambda}}
\newcommand{\Lgen}{\boldsymbol{\mathrm{L}}_{\Lmd}}
\label{sec: Framework}
\subsection{System}
We consider a quantum many-body system, whose Hamiltonian $H_V$ can be altered through an external field $V$. The system exchanges both energy and particles with a thermo-chemical reservoir, whose temperature $T$ and chemical potential $\mu$ we assume to be independently tunable. Hence, the dynamics of the system are controlled by three parameters which form the control vector $\Lmd=(V,T,\mu)\equiv(\Lambda^w,\Lambda^u,\Lambda^z)$.

A general thermodynamic cycle can be realised through periodic variation of these parameters, where the vector $\Lmd$ traces out a closed path $\gamma:t\rightarrow\Lmd_t$ in the space of control parameters. Once the system has settled into a periodic state, $\rho_t=\rho_{t+\tau}$, where $\tau$ denotes the period of the driving, the thermodynamic net effect of one cycle can be described in terms of the three variables
\begin{equation}
    X_\alpha=-\int_0^\tau dt \ f^\alpha_t\dot{\Lambda}^\alpha_t,\quad \text{where}\quad\alpha=w,u,z.
    \label{eq: X}
\end{equation}
Here, dots indicate derivatives with respect to time and the generalized forces,

\begin{equation}
    f^{w}_t=-\text{tr}\big[\rho_t\partial_{V}H_{V}\big]\Big|_{V=V_t}, \quad  f^{u}_t=-\text{tr}\big[\rho_t\log{\rho_t}\big],\quad \text{and}\quad f^{z}_t=\text{tr}\big[N\rho_t\big],
    \label{eq: therm f}
\end{equation}
correspond to the effective pressure, the entropy and the mean particle number of the system. Hence, $X_w$ is the applied work per cycle, and $X_u$ and $X_z$ correspond to the effective intake of thermal and chemical energy from the reservoir. Note that we set Boltzmann's constant to $1$ throughout.

The dissipated availability
\begin{equation}
    A=\sum_\alpha X_\alpha = -\int_0^\tau dt \ \sum_\alpha f^\alpha_t\dot{\Lambda}^\alpha_t
    \label{eq: A}
\end{equation}
provides a measure for the irreversibility of a thermodynamic process \cite{Salmon1983,Brandner2020}. To better understand the physical meaning of this quantity, we may rewrite it in terms of standard thermodynamic variables as follows. Upon integrating by parts and using the periodicity conditions, $f^\alpha_t=f^\alpha_{t+\tau}$ and $\Lmd_t=\Lmd_{t+\tau}$, we may express the dissipated availability as
\begin{align}
    A&=\int_0^\tau dt \ \big(\dot{S}_t T_t+\dot{N}_t\mu_t + P_t \big) = \int_0^\tau dt \ \big(\dot{S}_t T_t+\dot{N}_t\mu_t - J_t \big)\nonumber\\
    &=\int_0^\tau dt \ T_t\big(\dot{S}_t+\dot{Q}_t/T_t\big)=\int_0^\tau dt \ T_t \dot{\Sigma}_t\geq 0.
    \label{eq: A>=0}
\end{align}
Here, we have first identified the entropy $S_t=f^u_t$ and mean particle number $N_t=f^z_t$ of the system. In the second step, we have used the first law $\dot{E}_t=P_t+J_t$, where $E_t=\mathrm{tr}\big[H_{V_t}\rho_t\big]$ is the internal energy of the system, $P_t=f^w_t\dot{\Lambda}^w_t$ is the power applied by the mechanical driving and $J_t$ is the rate of energy uptake from the reservoir. Finally, we have introduced the heat flux into the reservoir $\dot{Q}_t=-\big(J_t-\mu_t\dot{N}_t\big)$ and the total rate of entropy production $\dot{\Sigma}_t=\dot{S}_t+\dot{Q}_t/T_t$. Since the second law requires $\dot{\Sigma}_t\geq0$, Eq.~\eqref{eq: A>=0} shows that $A\geq0$. 

\subsection{Quasi-static limit}
If the driving is quasi-static on the typical relaxation timescale of the system, its periodic state reduces to an instantaneous Gibbs state, i.e., $\rho_t=\varrho_{\Lmd_t}$ with 
\begin{equation}
    \varrho_{\Lmd}=\exp{\big[-(H_{V}-\mu N-\Phi_{\Lmd})/T\big]}\quad\text{and}\quad \Phi_{\Lmd}=-T\log{\big[\text{tr}\big[\exp{[-(H_{V}-\mu N)/T]}\big]\big]}.
    \label{eq: Gibbs state}
\end{equation}
The thermodynamic forces \eqref{eq: therm f} can then be expressed as partial derivatives of the grand-canonical potential $\Phi_{\Lmd}$ with respect to the conjugate control parameters. That is,
\begin{equation}
    f^{\alpha}_t\rightarrow\mathcal{F}^{\alpha}_{\Lmd_t}\quad\text{with}\quad\mathcal{F}^{\alpha}_{\Lmd}=-\partial_\alpha\Phi_{\Lmd},
\end{equation}
where $\partial_\alpha$ denotes the partial derivative with respect to $\Lambda^\alpha$. 

Since the generalized forces now depend parametrically on the driving protocol $\Lmd_t$, the thermodynamic quantities \eqref{eq: X} admit a geometric representation as line-integrals along the control path $\gamma$,
\begin{equation}
    \mathcal{X}_{\alpha}=-\int_0^\tau dt \ \mathcal{F}^{\alpha}_{\Lmd_t}\dot{\Lambda}^{\alpha}_t=\oint_{\gamma} \sum_{\beta}\mathcal{A}^{\alpha\beta}_{\Lmd}d\Lambda^{\beta}\quad\text{with}\quad\mathcal{A}^{\alpha\beta}_{\Lmd}=\Lambda^{\alpha}\partial_{\beta}\mathcal{F}^{\alpha}_{\Lmd}
    \label{eq: geom X}
\end{equation}
being the components of the thermodynamic vector potential, $\underline{\mathcal{A}}_{\Lmd}^{\alpha}=\big(\mathcal{A}^{\alpha w}_{\Lmd},\mathcal{A}^{\alpha u}_{\Lmd},\mathcal{A}^{\alpha z}_{\Lmd}\big)$. This result shows that the quantities $\mathcal{X}_\alpha$ do not depend on the parameterisation of the control path $\gamma$. A second geometric expression can be found by using Stokes' theorem to express the line-integral over the vector potentials as an integral over some surface $\boldsymbol{\Gamma}$ bounded by the path $\gamma$,
\begin{equation}
    \mathcal{X}_{\alpha}=\iint_{\boldsymbol{\Gamma}} \underline{\mathcal{B}}^{\alpha}_{\Lmd}\cdot d\boldsymbol{\Gamma}\quad\text{with}\quad\underline{\mathcal{B}}_{\Lmd}^{\alpha}=\nabla^{\Lmd}\times\underline{\mathcal{A}}_{\Lmd}^{\alpha}
    \label{eq: geom B}
\end{equation}
being the thermodynamic \emph{Berry curvature} and $\nabla^{\Lmd}=\big(\partial_{w},\partial_{u},\partial_{z}\big)$ denoting the gradient with respect to $\Lmd$. Finally, the dissipated availability becomes
\begin{equation}
    A=\sum_\alpha\mathcal{X}_\alpha=-\int_0^\tau dt \ \sum_{\alpha}\mathcal{F}^{\alpha}_{\Lmd_t}\dot{\Lambda}^{\alpha}_t=\int_{0}^{\tau} dt \ \dot{\Phi}_{\Lmd_t} = 0.
\end{equation}
Hence, any thermodynamic cycle is reversible in the quasi-static limit, as expected.

\subsection{Adiabatic response}
If the system is driven at finite speed but still slowly on its typical relaxation timescale, its state remains close to the instantaneous Gibbs state \eqref{eq: Gibbs state}. In this adiabatic-response (AR) regime, the generalized forces can be expanded to first order in the driving rates,
\begin{equation}
    f^{\alpha}_t=\mathcal{F}^{\alpha}_{\Lmd_t} + \sum_\beta R^{\alpha\beta}_{\Lmd_t}\dot{\Lambda}^{\beta}_t,
    \label{eq: gen f}
\end{equation}
where the adiabatic-response coefficients $R^{\alpha\beta}_{\Lmd_t}$ depend parametrically on the driving protocol. The dissipated availability thus becomes
\begin{equation}
    A=-\int_0^\tau dt \ \sum_{\alpha,\beta} R^{\alpha\beta}_{\Lmd_t}\dot{\Lambda}^{\alpha}_t\dot{\Lambda}^{\beta}_t =\int_0^\tau dt \sum_{\alpha,\beta}\ g^{\alpha\beta}_{\Lmd}\dot{\Lambda}^{\alpha}_t\dot{\Lambda}^{\beta}_t\geq0, \quad \text{where} \quad g^{\alpha\beta}_{\Lmd}=-(R^{\alpha\beta}_{\Lmd}+R^{\beta\alpha}_{\Lmd})/2,
    \label{eq:diss A AR}
\end{equation}
denotes the elements of a symmetric positive semi-definite matrix \cite{Brandner2020}, which can be interpreted as a pseudo-Riemannian metric in the space of control parameters. While $A$ is of second order in the driving rates, and thus not a geometric quantity, it still admits a geometric lower bound. Specifically, applying the Cauchy-Schwartz inequality gives
\begin{equation}
    A\geq\mathcal{L}^{2}/\tau, \quad \text{where} \quad \mathcal{L}\equiv\oint_\gamma \ \Bigg[\sum_{\alpha,\beta}g^{\alpha\beta}_{\Lmd}d\Lambda^\alpha d\Lambda^\beta\Bigg]^{1/2}
    \label{eq: A>L}
\end{equation}
denotes the \emph{thermodynamic length}, i.e., the length of the path $\gamma$ with respect to the metric $g^{\alpha\beta}_{\Lmd}$. This quantity provides a measure for the dissipation incurred by finite-time driving in the lowest order in $1/\tau$.

The inequality \eqref{eq: A>L} can be saturated for any control path $\gamma$ by optimizing its parameterisation. To this end, we make the thermodynamic length a functional of a monotonically increasing speed function $\phi_t$ by replacing $\Lmd_t$ with $\Lmd_{\phi_t}$. Solving the Euler-Lagrange equation for $\phi_t$ with respect to the boundary conditions $\phi_0=0$ and $\phi_\tau=\tau$ returns the optimal parameterisation which is implicitly given by the expression \cite{Brandner2020}
\begin{equation}
    t=\frac{1}{\mathcal{L}}\int_0^{\phi_t}ds \ \Bigg[\sum_{\alpha,\beta}g^{\alpha\beta}_{\Lmd_s}\dot{\Lambda}^{\alpha}_s\dot{\Lambda}^{\beta}_s\Bigg]^{1/2}.
    \label{eq: phit}
\end{equation}

\subsection{Systems with fixed particle number}
\label{sec: Systems with fixed particle number}
\newcommand{\lmd}{\boldsymbol{\lambda}}
\newcommand{\bM}{\bar{M}_{\lmd}}
\newcommand{\bR}{\bar{R}_{\lmd}}
\newcommand{\hF}[1]{\hat{\mathcal{F}}_{\lmd_{#1}}}
\newcommand{\hA}[1]{\hat{\mathcal{A}}_{\lmd_{#1}}}
\newcommand{\hg}[1]{\hat{g}_{\lmd_{#1}}}
\newcommand{\hR}[1]{\hat{R}_{\lmd_{#1}}}
\newcommand{\hB}[1]{\hat{\mathcal{B}}_{\lmd_{#1}}}
Our analysis thus far has assumed a grand-canonical setting, where the system has access to a thermo-chemical reservoir. However, in the context of thermal machines, one is usually interested in thermodynamic cycles, where the system exchanges only heat with its environment, while its particle number remains constant. Such cycles must generally be described in a canonical setting, which is however technically more difficult to implement than the grand-canonical one. In the quasi-static limit, this problem can be circumvented by applying the principle of equivalence of ensembles. That is, for sufficiently large systems, equilibrium quantities can be calculated in the grand-canonical ensemble with the mean particle number being fixed afterwards by tuning the chemical potential. This approach should still be valid in the AR regime, where the system remains close to equilibrium. Under this assumption, the grand-canonical framework of thermodynamic geometry can be systematically and consistently reduced to a canonical one, as we show in the following.

To fix the particle number, we have to choose the chemical potential $\Lambda^z_t$ such that
\begin{equation}
    f^z_t=\mathcal{N}
\end{equation}
holds for any $0\leq t\leq\tau$ and for some value $\mathcal{N}\gg1$. In AR, this condition yields the relation
\begin{equation}
    \mathcal{F}^{z}_{\Lmd_t}+\sum_{\alpha}R^{z\alpha}_{\Lmd_t}\dot{\Lambda}^{\alpha}_t=\mathcal{N},
    \label{eq: Fz AR}
\end{equation}
which can be perturbatively solved for $\Lambda^{z}_t$ using the ansatz
\begin{equation}
    \Lambda^z_t=\Lambda^{z0}_{\lmd_t}+\sum_{a}\Lambda^{za}_{\lmd_t}\dot{\lambda}_t^{a}
    \label{eq: Lamz ansatz}
\end{equation}
with yet undetermined coefficients $\Lambda^{za}_{\lmd}$. Here, we have introduced the reduced control vector $\lmd=(\Lambda^w,\Lambda^u)\equiv(\lambda^w,\lambda^u)$. Furthermore, we use the convention that Greek indices run over $w,u,z$ and Latin ones run over $w,u$. 

Inserting the ansatz \eqref{eq: Lamz ansatz} into Eq.~\eqref{eq: Fz AR} and expanding to first order in the control rates gives
\begin{align}
    &\mathcal{F}^z_{\lmd,\Lambda^{z0}}+M^{zz}_{\lmd,\Lambda^{z0}}\sum_{a}\Lambda^{za}_{\lmd}\dot{\lambda}^{a}+R^{zz}_{\lmd,\Lambda^{z0}}\sum_{a}\big(\partial_a\Lambda^{z0}_{\lmd}\big)\dot{\lambda}^{a}+\sum_{a}R^{za}_{\lmd,\Lambda^{z0}}\dot{\lambda}^a\nonumber\\
    = \ &\mathcal{F}^z_{\lmd,\Lambda^{z0}}+\sum_{a}\Big(M^{zz}_{\lmd,\Lambda^{z0}}\Lambda^{za}_{\lmd}+R^{zz}_{\lmd,\Lambda^{z0}}\big(\partial_a\Lambda^{z0}_{\lmd}\big)+R^{za}_{\lmd,\Lambda^{z0}}\Big)\dot{\lambda}^a=\mathcal{N},
    \label{eq: Fz exp AR}
\end{align}
where we have defined the coefficients 
\begin{equation}
    M^{\alpha\beta}_{\Lmd}=\partial_{\alpha}\mathcal{F}^{\beta}_{\Lmd}=-\partial_{\alpha}\partial_{\beta}\Phi_{\Lmd}= M^{\beta\alpha}_{\Lmd}
\end{equation}
and $\partial_{a}$ denotes a partial derivative with respect to $\lambda_a$.
We note that we drop all time arguments until Eq.~\eqref{eq: qs forces} for notational clarity. In zeroth order, Eq.~\eqref{eq: Fz exp AR} yields the condition
\begin{equation}
    \mathcal{F}^{z}_{\lmd,\Lambda^{z0}}=\mathcal{N},
    \label{eq: Fz zeroth order}
\end{equation}
which makes $\Lambda^{z0}$ a function of $\lmd$; in practice, the function $\Lambda^{z0}_{\lmd}$ will typically have to be found numerically. Furthermore, taking a time derivative of Eq.~\eqref{eq: Fz zeroth order} gives
\begin{equation}
    \sum_{a}\Big(\bar{M}^{az}_{\lmd}+\bar{M}^{zz}_{\lmd}\big(\partial_a\Lambda^{z0}_{\lmd}\big)\Big)\dot{\lambda}^a=0,
    \label{eq: summands}
\end{equation}
with $\bM^{\alpha\beta}=M^{\alpha\beta}_{\Lmd}\big|_{\Lambda^z=\Lambda^{z0}_{\lmd}}$. Since the parameters $\lambda^a$ are independently controlled, the summands in Eq.~\eqref{eq: summands} have to vanish individually. We thus have
\begin{equation}
    \partial_a \Lambda^{z0}_{\lmd}=-\bM^{az}\big/\bM^{zz}.
    \label{eq: Lamz0 result}
\end{equation}
Together, with this result and the condition \eqref{eq: Fz zeroth order}, Eq.~\eqref{eq: Fz exp AR} implies
\begin{equation}
    \Lambda^{za}_{\lmd}=\Big(-\bR^{za}+\bR^{zz}\bM^{az}\big/\bM^{zz}\Big)\big/\bM^{zz}
    \label{eq: Lamz1 result}
\end{equation}
with $\bR^{\alpha\beta}=R^{\alpha\beta}_{\Lmd}\big|_{\Lambda^{z}=\Lambda^{z0}_{\lmd}}$. The relations \eqref{eq: Lamz0 result} and \eqref{eq: Lamz1 result} fully determine the chemical potential $\Lambda^z_t$ for any $\lmd_t$ at all times $0\leq t \leq\tau$, up to second-order corrections in the driving rates. 

We can now construct a geometric framework for thermodynamic cycles with fixed particle number. First, in the quasi-static limit, the generalized forces are given by
\begin{equation}
    f^a_t=\hF{t}^a\quad \text{with}\quad\hF{}^a = \mathcal{F}_{\Lmd}^a \Big|_{\Lambda^z=\Lambda^{z0}_{\lmd}}.
    \label{eq: qs forces}
\end{equation}
Our two main quantities of interest, the applied work $\mathcal{X}_w$ and the uptake of thermal energy $\mathcal{X}_u$, thus become
\begin{equation}
    \mathcal{X}_a=-\int_0^\tau dt \ \hF{t}^a\dot{\lambda}^a_t=\oint_{\hat{\gamma}} \sum_{b}\hat{\mathcal{A}}^{ab}_{\lmd}\cdot d\lambda^{b}=\iint_{\hat{\Gamma}}\hB{}^a d\hat{\Gamma}
    \label{eq: canon Xa}
\end{equation}
with
\begin{equation}
    \hA{}^{ab}=\lambda^a\partial_{b}\hF{}^{a}=\lambda^a\Big(\bM^{ba}+\bM^{za}\partial_{b}\Lambda^{z0}_{\lmd}\Big)=\lambda^a\Big(\bM^{ba}-\bM^{za}\bM^{bz}\big/\bM^{zz}\Big)
    \label{eq:vecA canon}
\end{equation}
being the components of the reduced thermodynamic vector potential and the reduced Berry curvatures being given by
\begin{equation}
    \hB{}^{a}=\partial_w \hA{}^{au}-\partial_u \hA{}^{aw}=\sum_{b}\epsilon_{ab}\frac{\hA{}^{ab}}{\lambda^{a}}+\frac{\lambda^{a}}{\bM^{zz}}\sum_{b,c}\epsilon_{bc}\big[\bM^{zbz}\bM^{cz}+\bM^{cza}\bM^{bz}\big]
    \label{eq: geom hB}
\end{equation}
where $M^{\alpha\beta\gamma}_{\Lmd}=\partial_{\alpha}M^{\beta\gamma}_{\Lmd}$ and $\bM^{\alpha\beta\gamma}=M^{\alpha\beta\gamma}_{\Lmd}\big|_{\Lambda^z=\Lambda^{z0}}$ and $\epsilon_{ab}=\delta_{wa}\delta_{ub}-\delta_{ua}\delta_{wb}$.

Here, $\hat{\gamma}$ is the reduced control path in the $w$-$u$ plane and $\hat{\Gamma}$ is the area it encircles. In AR, the generalized forces are connected to the driving rates by the relations
\begin{equation}
    f^a_t=\hF{t}^{a}+\sum_b \hR{t}^{ab}\dot{\lambda}^b_t,
    \label{eq: fa}
\end{equation}
with the reduced adiabatic-response coefficients
\begin{align}
    \hR{}^{ab}&\equiv\bR^{ab}+\bM^{za}\Lambda^{zb}+\bR^{az}\Big(\partial_a\Lambda^{z0}_{\lmd}\Big)\nonumber\\
    &=\bR^{ab}-\bM^{za}\bR^{zb}\big/\bM^{zz}-\bM^{bz}\bR^{za}\big/\bM^{zz}+\bM^{za}\bM^{bz}\bR^{zz}\big/\big(\bM^{zz}\big)^2,
    \label{eq: Reduced R}
\end{align}
which can be found by inserting the ansatz \eqref{eq: Lamz ansatz} into the expression for the generalized forces Eq.~\eqref{eq: gen f}, expanding to first order in the driving rates $\dot{\lambda}^a$, and substituting in the relations \eqref{eq: Lamz0 result} and \eqref{eq: Lamz1 result}. 

To verify that our approach is thermodynamically consistent we still have to show the second law is upheld. To this end, we consider the dissipated availability $A$ for a general thermodynamic cycle, insert Eq.~\eqref{eq: Lamz ansatz} for the chemical potential $\Lambda^z_t$ and expand to second order in the driving rates. These steps yield
\begin{equation}
    A=\int_0^\tau dt \ \sum_{\alpha,\beta} R^{\alpha\beta}_{\Lmd}\dot{\Lambda}^{\alpha}_t\dot{\Lambda}^{\beta}_t=\int_0^\tau dt \ \sum_{a,b}\hR{t}^{ab}\dot{\lambda}^{a}_t\dot{\lambda}^{b}_t=\int_0^\tau dt \ \sum_{a,b}\hg{t}^{ab}\dot{\lambda}^{a}_t\dot{\lambda}^{b}_t\equiv\hat{A}
\end{equation}
with $\hg{}^{ab}=-\big(\hR{}^{ab}+\hR{}^{ba}\big)/2$. Hence, the reduced dissipated availability $\hat{A}$ can be expressed in the same form as $A$ with $g^{\alpha\beta}_{\Lmd}$ replaced by $\hg{}^{ab}$. It remains to show that the coefficients $\hg{}^{ab}$ define a positive semi-definite matrix, which can be interpreted as a pseudo-Riemannian metric in the reduced space of control parameters. To this end, we observe that, for arbitrary $X^a,X^b\in\mathbb{R}$, 
\begin{equation}
    \sum_{a,b}\hg{}^{ab}X^aX^b=\sum_{\alpha,\beta} g^{\alpha\beta}_{\Lmd}X^\alpha X^\beta\big|_{\Lambda^{z}\rightarrow\Lambda^{z0}_{\lmd}}\quad\text{with}\quad X^z=-\big(\bM^{zw}X^w + \bM^{zu}X^{u}\big)\big/\bM^{zz}.
\end{equation}
Since the elements $g^{\alpha\beta}_{\Lmd}$ form a positive semi-definite matrix for any $\Lmd$, it follows that $\sum_{a,b}\hg{}^{ab}X^aX^b\geq0$, which is the desired result.

With these prerequisites, it is now straightforward to derive a geometric bound on $\hat{A}$. From the Cauchy-Schwartz inequality, we have
\begin{equation}
    \hat{A}\geq\hat{\mathcal{L}}^2/\tau\quad\text{with}\quad\hat{\mathcal{L}}=\oint_{\hat{\gamma}} \Bigg[\sum_{a,b}\hg{}^{ab}d\lambda^{a}d\lambda^{b}\Bigg]^{1/2}
    \label{eq: canon L}
\end{equation}
being the reduced thermodynamic length. As before, this bound can be saturated for any reduced control path $\hat{\gamma}$ by optimizing its parameterization. The corresponding optimal speed function is implicitly defined by the condition
\begin{equation}
    t=\frac{1}{\hat{\mathcal{L}}}\int_0^{\hat{\phi}_t}ds \ \Bigg[\sum_{a,b}\hg{}^{ab}d\lambda^{a}d\lambda^{b}\Bigg]^{1/2}.
    \label{eq: canon sf}
\end{equation}

\section{Many-body heat engines}
\label{sec: MBHE}
In the following, we show how the general framework developed in the previous section makes it possible to describe heat engines that use an ideal quantum gas as a working system. As a key application of this theory, we demonstrate how the enhancement of engine performance through Bose-Einstein condensation, which has been described recently in Ref.~\cite{Myres2022}, can be understood through the lens of thermodynamic geometry.

\subsection{Geometry of engine cycles}
\label{sec: Thermodynamics of heat engines}
We begin with a brief review of the thermodynamic geometry of heat engines that are driven by continuous periodic variations of a mechanical control parameter, $V=\lambda^w$, and the temperature of a thermal reservoir, $T=\lambda^u$; for details see \cite{Brandner2020}. Note that, here, we consider the canonical setting, where the working system exchanges only thermal energy with its environment.

With the definitions of the previous section, output and input of a general engine cycle are given by
\begin{equation}
    W=-X_w=\int_0^\tau dt \ f^w_t \dot{\lambda}^w_t\quad\text{and}\quad
    U=X_u=-\int_0^\tau dt \ f^u_t \dot{\lambda}^u_t,
    \label{eq: engine W U}
\end{equation}
where $W$ is the generated work per cycle and $U$ is the thermal energy uptake from the reservoir. The dissipated availability per cycle is thus given by
\begin{equation}
    A=U-W\geq0,
    \label{eq: engine A}
\end{equation}
which shows that $U$ can be understood as the available energy for work production. It is therefore natural to define the efficiency of the engine as 
\begin{equation}
    \varepsilon=W/U\leq1,
    \label{eq: gen eff}
\end{equation}
where the driving protocol $\lmd_t$ must be chosen such that $W>0$.

A second figure of merit for a heat engine is given by its power output $P=W/\tau$. In AR, this figure is linked to the efficiency \eqref{eq: gen eff} by the universal trade-off relation
\begin{equation}
    (1-\varepsilon)\big(\mathcal{W}/\mathcal{L}\big)^2\equiv(1-\varepsilon)\Psi\geq\mathcal{W}\big/\tau=P,
    \label{eq: trade-off}
\end{equation}
where $\mathcal{W}=-\mathcal{X}_w$ is the geometric work and $\mathcal{L}$ is the thermodynamic length of the control path that is mapped out by the driving protocol $\lmd_t$. Notably, equality is achieved in Eq.~\eqref{eq: trade-off} if $A=\mathcal{L}^2/\tau$. That is, the trade-off relation can be saturated for any control path by optimizing its parameterization so that it minimizes the dissipated availability. 

The trade-off relation \eqref{eq: trade-off} shows that the power of a generic heat engine decays at least linearly to 0 as its efficiency approaches its maximum value in the quasi-static limit $\tau\rightarrow\infty$, where $A\rightarrow0$. The slope of this decay is determined by a purely geometric figure of merit $\Psi\equiv\big(\mathcal{W}/\mathcal{L}\big)^2$. Although derived originally for microscopic heat engines, these results do not rely on any assumption on the size of system and hence are directly applicable to many-body engines. The figures $\mathcal{W}$ and $\Psi$ can then be used to quantitatively probe the role of collective effects for the performance of general engine cycles in AR.

\subsection{BEC engine}
\newcommand{\adn}{\boldsymbol{\mathrm{a}}}
\newcommand{\ad}{\boldsymbol{\mathrm{a}}^{\dagger}}
\newcommand{\Ad}{\mathrm{A}^{\dagger}_{\boldsymbol{\mathrm{n}}}}
\newcommand{\Adm}{\mathrm{A}^{\dagger}_{\boldsymbol{\mathrm{m}}}}
\newcommand{\A}{\mathrm{A}_{\boldsymbol{\mathrm{n}}}}
\newcommand{\Adt}{\mathrm{A}^{\dagger}_{\boldsymbol{\mathrm{n}}}}
\newcommand{\At}{\mathrm{A}_{\boldsymbol{\mathrm{n}}}}
\newcommand{\bu}[1]{\boldsymbol{\mathrm{#1}}}
\newcommand{\Li}[2]{\mathrm{Li}_{#1}(#2)}
To further develop this idea, we now consider a many-body heat engine, whose working system consists of an ideal Bose gas in a two-dimensional, isotropic harmonic trap with tunable strength $\omega=V$. This system is capable of hosting one of the simplest collective quantum effects, the formation of an ideal Bose-Einstein condensate, which has recently been shown to enhance engine performance under certain conditions \cite{Myres2022}. To keep this paper self-contained, we first present the relevant background material describing the equilibrium thermodynamics of the system \cite{Ketterle1996,pitaevskii2003}. We then lay down a dynamical model based on a quantum master equation, which makes it possible to calculate the thermodynamic length $\mathcal{L}$, thus giving us access to the figure of merit $\Psi$. In both cases, we start with a grand-canonical description, which is technically easier to implement than a canonical one, and then fix the particle number in the working system by tuning the chemical potential of the reservoir.

\subsubsection{Equilibrium}\
\label{sec: equilibrium BEC}

\noindent We consider a collection of non-interacting bosons with mass $m$ confined by a 2-dimensional, harmonic trap with strength $\omega$. The single particle Hamiltonian reads
\begin{equation}
    H^{1}_{\omega}=\hbar\omega\big(\ad\adn+1\big),
    \label{eq: H1}
\end{equation}
with the usual creation and annihilation operators $\adn\equiv\sqrt{m\omega/2}(\boldsymbol{\mathrm{x}}+i\boldsymbol{\mathrm{p}}/m\omega)$, $\ad\equiv\sqrt{m\omega/2}(\boldsymbol{\mathrm{x}}-i\boldsymbol{\mathrm{p}}/m\omega)$ and the 2-dimensional position and momentum operators $\boldsymbol{\mathrm{x}}$ and $\boldsymbol{\mathrm{p}}$. The Hamiltonian is readily extended to a many-body system by introducing the Fock-space operators, $\A$ and $\Ad$, which remove and add particles to the single-particle eigenstates $|\boldsymbol{\mathrm{n}}\rangle$. Both Fock-space operators are implicitly dependent on the trap strength.
According to the standard rules of second quantization, we have
\begin{equation}
    H_\omega=\sum_{\mathrm{m},\mathrm{n}}\langle\bu{m}|H^1_\omega|\bu{n}\rangle\Adm\A=\sum_{\mathrm{n}}E_{\bu{n}}\Ad\A,
    \label{eq: Fock H}
\end{equation}
where $\bu{n}=(n_x, n_y)\in\mathbb{N}^2_0$ and $E_{\bu{n}}=\hbar\omega(n_x+n_y+1)$ are the single-particle energy levels. 

In the grand-canonical ensemble, the thermostatics of the system are fully determined by the grand potential,
\begin{equation}
    \Phi_{\Lmd}=T\sum_{n_x,n_y=0}^{\infty}\log{\big[1-q^{n_x+n_y}z\big]},
\end{equation}
where $q=\exp[-\hbar\omega/T]<1$ and $z=\exp[(\mu-\hbar\omega )/T]\leq1$. Upon using the series expansions $\log{[1-x]}=-\sum_{j=1}^{\infty}x^{j}/j$ and $[1-x]^{-1}=\sum_{j=0}^{\infty}x^j$ for $x<1$, this expression can be rewritten as
\begin{align}
    \Phi_{\Lmd}&=-T\sum_{j=1}^{\infty}\frac{z^{j}}{j}\sum_{n_x,n_y=0}^{\infty}q^{(n_x+n_y)j}=-T\sum_{j=1}^{\infty}\frac{z^{j}}{j}\sum_{n=0}^{\infty}(n+1)q^{nj}\nonumber\\
    &=-T\Bigg(\sum_{j=1}^{\infty}\frac{z^{j}}{j}+\sum_{j=1}^{\infty}\frac{z^{j}}{j}\frac{2q^j-q^{2j}}{(1-q^{j})^{2}}\Bigg).
    \label{eq: BEC Phi}
\end{align}

Up to this point, no approximations have been made. To obtain a tractable expression for $\Phi_{\Lmd}$, we now make the assumption $\hbar\omega/T\ll1$ and expand the denominator of the second term in \eqref{eq: BEC Phi} to leading order in $T/\hbar\omega$, which yields \cite{Ketterle1996}
\begin{equation}
    \Phi_{\Lmd}=-T\Bigg(\Li{1}{z}+\frac{2\Li{2}{zq}-\Li{2}{zq^{2}}}{(\hbar\omega/T)}+\frac{2\Li{3}{zq}-\Li{3}{zq^{2}}}{(\hbar\omega/T)^2}\Bigg).
    \label{eq: phi BEC 2d}
\end{equation}
Here, we have used the definition of the polylogarithmic functions, $\Li{s}{z}=\sum_{k=1}^{\infty}z^{k}/k^{s}$. The average particle number of the system is now given by
\begin{align}
    \mathcal{N}&=-\partial_{\mu}\Phi_{\Lmd}=\mathcal{F}^{z}_{\Lmd}\nonumber\\
    &\simeq\Li{0}{z}+\frac{2\Li{1}{zq}-\Li{1}{zq^{2}}}{(\hbar\omega/T)}+\frac{2\Li{2}{zq}-\Li{2}{zq^{2}}}{(\hbar\omega/T)^2}\equiv\mathcal{N}_0+\mathcal{N}'.
    \label{eq: N BEC fix}
\end{align} 
The first term in Eq.~\eqref{eq: N BEC fix} corresponds to the groundstate population $\mathcal{N}_0$ and the sum of the additional terms to the population of excited states, $\mathcal{N}'$. 

We can now transition to an effectively canonical picture by using Eq.~\eqref{eq: N BEC fix} to fix the chemical potential $\mu=\Lambda^{z}$, or equivalently the effective fugacity $z$, for a given $\mathcal{N}$. Since $\mathcal{N}'$ is bounded by $\mathcal{N}'_{\text{max}}=\mathcal{N}'\big|_{z=1}$, it follows that the ground state must be macroscopically occupied for $\mathcal{N}>\mathcal{N}'_{\text{max}}$. Upon evaluating $\mathcal{N}'_{\text{max}}$ to leading order in $\hbar\omega/T$, this condition yields the critical temperature
\begin{equation}
    T_c\simeq\hbar\omega\frac{\sqrt{6\mathcal{N}}}{\pi}.
    \label{eq: Tc}
\end{equation}
For $T<T_c$, the system hosts a Bose-Einstein condensate. In the following we refer to this regime as the condensate phase and the regime $T> T_c$ as the normal phase.

\subsubsection{Lindblad dynamics}\

\noindent To describe the dynamics of the system in AR, we can formulate a quantum master equation. To this end, we assume that the system is weakly coupled to a thermo-chemical reservoir with temperature $T$ and chemical potential $\mu$, where $\omega$, $T$ and $\mu$ are independently tunable. Provided that the control parameters vary slowly on the relaxation timescale of the system, the time evolution of its state $\rho_t$ can be thermodynamically consistently modelled with the Lindblad equation \cite{Alicki1979}
\begin{equation}
    \dot{\rho}_t=\boldsymbol{\mathrm{L}}_t\rho_t, \quad\text{where}\quad\boldsymbol{\mathrm{L}}_t=-\frac{i}{\hbar}\big[H_{\omega_t},\rho_t\big] + \boldsymbol{\mathrm{D}}_{\Lmd_t}.
    \label{eq: Lin ME}
\end{equation}
The first term of $\boldsymbol{\mathrm{L}}_t$ accounts for the unitary evolution of the system and the dissipation super-operator
\begin{equation}
    \boldsymbol{\mathrm{D}}_{\Lmd}\boldsymbol{\cdot}\equiv\kappa\sum_{\bu{n}}\mathfrak{n}_{\bu{n},\Lmd}\bigg(\At\boldsymbol{\cdot}\Adt-\frac{1}{2}\big\{\boldsymbol{\cdot},\Adt\At\big\}\bigg)+\big(\mathfrak{n}_{\bu{n},\Lmd}+1\big)\bigg(\Adt\boldsymbol{\cdot}\At-\frac{1}{2}\big\{\boldsymbol{\cdot},\At\Adt\big\}\bigg)
\end{equation}
describes the exchange of particles with the reservoir. Here,
\begin{equation}
    \mathfrak{n}_{\bu{n},\Lmd}=1/\big(1-\exp[-(E_{\bu{n}}-\mu)/T]\big)
\end{equation}
is the Bose-Einstein factor and $\kappa$ is the rate setting the relaxation timescale. 

The master equation \eqref{eq: Lin ME} has the instantaneous stationary solution $\varrho_{\Lmd}=\exp[-(H_\omega-N\mu)/T]$ with $N=\sum_{\bu{n}}\Ad\A$. That is, if the control parameters are fixed, the system relaxes to a Gibbs state. This condition is both sufficient and necessary for thermodynamic consistency in the weak-coupling limit \cite{Spohn1978b,Spohn1978,Brandner2016}. Furthermore, reflecting micro-reversibility, the dissipation super-operator obeys the quantum detailed balance condition
\begin{equation}
    \boldsymbol{\mathrm{D}}_{\Lmd}\varrho_{\Lmd}= \varrho_{\Lmd}\boldsymbol{\mathrm{D}}_{\Lmd}^{\dagger},
\end{equation}
where $\boldsymbol{\mathrm{D}}_{\Lmd}^{\dagger}$ denotes the adjoint dissipator,
\begin{equation}
    \boldsymbol{\mathrm{D}}_{\Lmd}^{\dagger}\boldsymbol{\cdot}\equiv\kappa\sum_{\bu{n}}\big(\mathfrak{n}_{\bu{n},\Lmd}+1\big)\bigg(\Adt\boldsymbol{\cdot}\At-\frac{1}{2}\big\{\boldsymbol{\cdot},\Adt\At\big\}\bigg)+\mathfrak{n}_{\bu{n},\Lmd}\bigg(\At\boldsymbol{\cdot}\Adt-\frac{1}{2}\big\{\boldsymbol{\cdot},\At\Adt\big\}\bigg).
\end{equation}

\subsubsection{Stirling cycle} \

\begin{figure}
    \centering
    \includegraphics[width=0.8\textwidth]{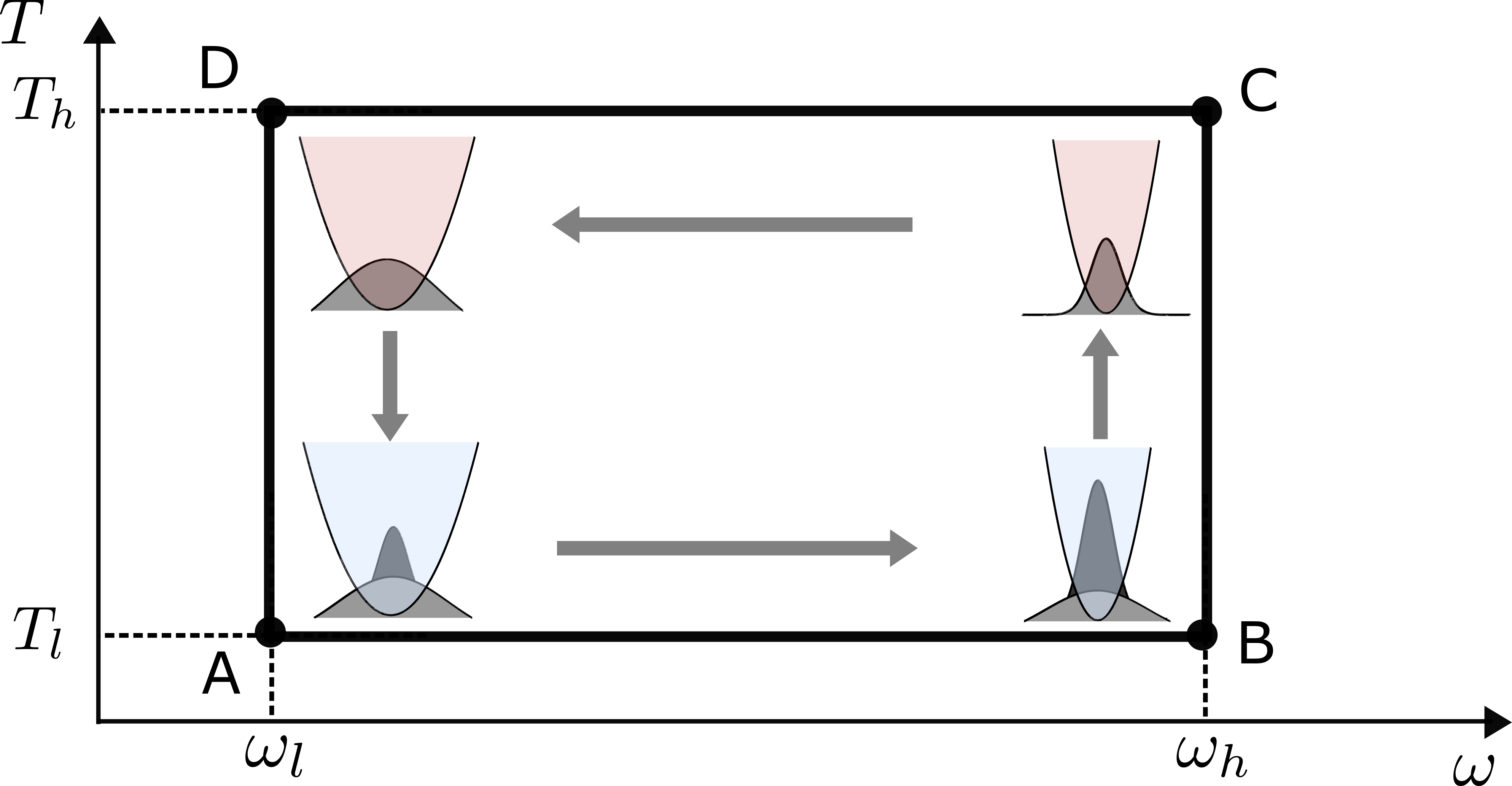}
    \caption{Outline of a typical Stirling cycle with a harmonic trap and a schematic of the population distribution of the BEC population (dark-grey) and excited states population (light-grey). Starting at $\mathrm{A}$, in the first stroke, $\mathrm{A}\rightarrow\mathrm{B}$, the system is operated at the cooler temperature, $T_l$, whilst the trap frequency is increased, $\omega_l\rightarrow\omega_h$. In this `compression' stroke, work is done on the system. In $\mathrm{B}\rightarrow\mathrm{C}$ the trap strength is kept fixed at $\omega_h$ and the temperature is increased $T_l\rightarrow T_h$, adding heat to the system. The trap strength is then returned, keeping $T_h$ fixed, in the `expansion' stroke $\mathrm{C}\rightarrow\mathrm{D}$, decreasing the level spacing between energy levels and extracting work from the system. Finally in $\mathrm{D}\rightarrow\mathrm{A}$, the temperature is returned to $T_l$. The cyclic modulation of the parameters in $\lmd$ maps the path $\hat{\gamma}$ in the $\omega$-$T$ plane.}
    \label{fig: Stirling schematic}
\end{figure}

\noindent For concreteness, we now focus on a Stirling cycle which consists of four distinct strokes: two isothermal ones, where the temperature $T$ is fixed and the trap frequency, which plays the role of the inverse volume, is varied between the initial and final values $\omega_{l}$ and $\omega_{h}=\omega_l+\Delta\omega$ with $\Delta\omega>0$, and two isochoric strokes, where the trap frequency is fixed and the temperature is varied between $T_l$ and $T_h=T_l+\Delta T$ with $\Delta T>0$. An outline of this cycle is shown in Fig.~\ref{fig: Stirling schematic}. 

To apply the general framework developed in Sec.~\ref{sec: Framework}, we require two types of quantities. First, the quasi-static generalized forces, $\mathcal{F}^{\alpha}_{\Lmd}=-\partial_{\alpha}\Phi_{\Lmd}$, which can be readily obtained from the expression \eqref{eq: phi BEC 2d} for the grand-canonical potential, are given by
\begin{align}
    \mathcal{F}_{\Lmd}^w=&-\hbar\bigg(\Li{0}{z}+\frac{4\Li{1}{zq}-3\Li{1}{zq^2}}{\big(\hbar\omega/T\big)}+\frac{6\Li{2}{zq}-4\Li{2}{zq^2}}{\big(\hbar\omega/T\big)^2}\nonumber\\
    &+\frac{4\Li{2}{zq}-2\Li{2}{zq^2}}{\big(\hbar\omega/T\big)^3}\bigg),\label{eq: Fw BEC}\\
    \mathcal{F}_{\Lmd}^u=\ &\frac{1}{T}\bigg(\Phi_{\Lmd}+\xi_{1}\Li{0}{z}+\frac{\xi_22\Li{1}{zq}-\xi_3\Li{1}{zq^2}}{\big(\hbar\omega/T\big)}+\frac{2\xi_2\Li{2}{zq}-\xi_3\Li{2}{zq^2}}{\big(\hbar\omega/T\big)^2}\bigg)\nonumber\\
    &+\frac{2\Li{2}{zq}-\Li{2}{zq^2}}{\big(\hbar\omega/T\big)}+\frac{4\Li{2}{zq}-2\Li{2}{zq^2}}{\big(\hbar\omega/T\big)^2},\\
    \mathcal{F}_{\Lmd}^z=&\ \Li{0}{z}+\frac{2\Li{1}{zq}-\Li{1}{zq^{2}}}{(\hbar\omega/T)}+\frac{2\Li{2}{zq}-\Li{2}{zq^{2}}}{(\hbar\omega/T)^2},
    \label{eq: Fz BEC}
\end{align}
where $\xi_n=n\hbar\omega-\mu$. Second, solving the master equation \eqref{eq: Lin ME} to first order in adiabatic perturbation theory gives the periodic state of the system in the form $\rho_t\simeq\varrho_{\Lmd_t}+\sum_{\alpha}\varrho^{\alpha}_{\Lmd_t}\dot{\Lambda}^{\alpha}_t$ \cite{Cavina2017}. Inserting this state into the definitions of the generalized forces \eqref{eq: gen f} and expanding to first order in the driving rates yields the primary AR coefficients
\begin{equation}
    R^{\alpha\beta}_{\Lmd}=-\frac{1}{\kappa}\bigg(\frac{1}{T}\sum_{n=0}^{\infty}g_n D^{\alpha}_{n,\Lmd}D^{\beta}_{n,\Lmd}\mathfrak{n}_{n,\Lmd}\big(\mathfrak{n}_{n,\Lmd}+1\big)+\delta_{ww,\alpha\beta}\sum_{n=0}^{\infty}C_{\Lmd}h_n \mathfrak{n}_{n+2,\Lmd}\big(\mathfrak{n}_{n,\Lmd}+1\big)\bigg),
    \label{eq: AR coeff BEC}
\end{equation}
with coefficients $D^{\alpha}_{n}$ and $C_{n}$ defined as
\begin{equation}
    D^{z}_{n,\Lmd}=1,\quad D^{u}_{n,\Lmd}=\frac{(E_{n}-\mu)}{T},\quad D^{w}_{n,\Lmd}=-\hbar(n+1)\quad\text{and}\quad C_{\Lmd}=\frac{\hbar(\exp[2\hbar\omega/T]-1)}{4\omega},
\end{equation}
and degeneracies $g_n=(n+1)$ and $h_n=\frac{2}{3}(n+1)(n+2)(n+3)$. The approximate analytic forms of these coefficients are found in the same way as with the grand potential, see \ref{sec: Appendix D}.

To carry out the reduction to an effectively canonical system, to which the trade-off relation \eqref{eq: trade-off} applies, we still need the coefficients $M^{\alpha\beta}_{\Lmd}=\partial_{\alpha}\mathcal{F}^{\beta}_{\Lmd}$ and $M^{\alpha\beta\gamma}_{\Lmd}=\partial_{\alpha}M^{\beta\gamma}_{\Lmd}$. These quantities can be directly obtained from the expressions \eqref{eq: Fw BEC}-\eqref{eq: Fz BEC} for the quasi-static generalized forces and are spelled out in \ref{sec: Appendix B} for reference. The reduced thermodynamic Berry curvature $\hB{}^{a}$ and AR coefficients $\hat{R}^{ab}_{\lmd}$ can thus be constructed from Eqs.~\eqref{eq: geom hB} and \eqref{eq: Reduced R} upon eliminating $\Lambda^z$ by numerically solving the constraint $\mathcal{F}^z_{\Lmd}=\mathcal{N}$ for $z$. With these prerequisites, all relevant thermodynamic quantities can now be calculated.

\subsubsection{Results} \ 

\noindent We consider Stirling cycles with $\Delta\omega=0.1\omega_l$ and $\Delta T=50\hbar\omega_l$ for a range of $T_l$, from $T_l=10\hbar\omega_l$ to $2.5\times10^3\hbar\omega_l$. Each Stirling cycle is performed for a range of systems with fixed average particle number from $\mathcal{N}=10^{3}$ to $10^{6}$. The critical temperature is given by Eq.~\eqref{eq: Tc}.
We identify three distinct regimes of operation. First, for $T_l\ll T_c$ the engine operates only inside the condensate phase. In this regime, the net work output is suppressed by the presence of a BEC, which reduces the effective pressure of the gas and thus the work that can be extracted during the expansion stroke. Second, for $T_l\gg T_c$, the engine operates completely in the normal phase, where its performance resembles that of a collection of $\mathcal{N}$ independent single-particle engines (SPE) as the quantum gas behaves increasingly classical. Finally, if $T_l\sim T_c$, the cycle crosses the transition between the two phases. In this regime, the compression stroke $(\mathrm{A}\rightarrow\mathrm{B})$ and expansion stroke $(\mathrm{C}\rightarrow\mathrm{D})$ are performed with largely different groundstate populations. While the expansion occurs in the vapour-phase with normal effective pressure, the compression is carried out in the presence of a BEC, which substantially reduces the effective pressure and thus the amount of work required to increase the trap strength. This effect, which has been described before in Ref.~\cite{Myres2022}, leads to a significant increase in the net extracted work compared to cycles operating solely in the vapour- or condensate phase. 

\begin{figure}
    \centering
    \includegraphics[width=\textwidth]{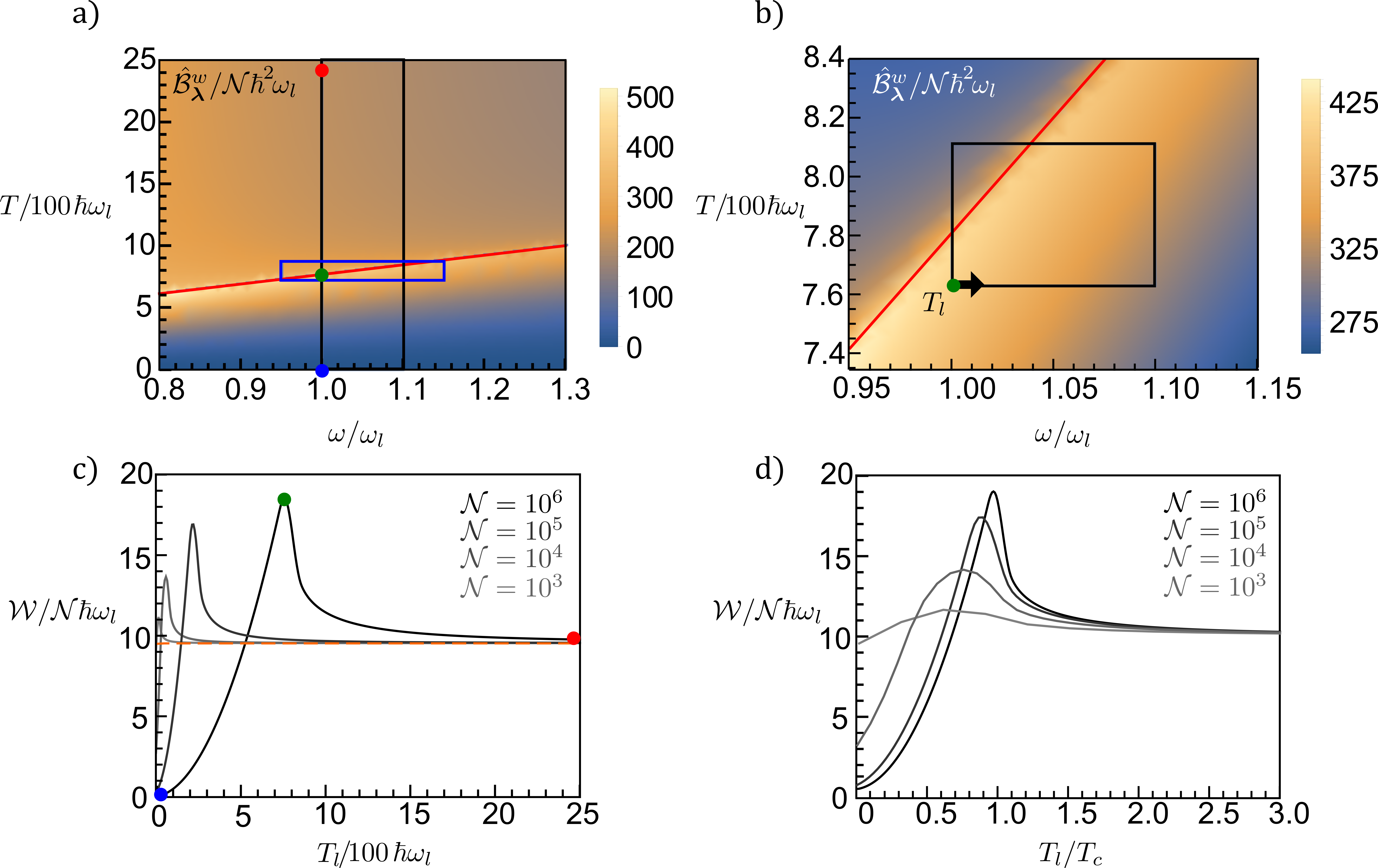}
    \caption{Quasi-static results. a) Plot of the normalised Berry curvature $\hB{}/\mathcal{N}$, for $\mathcal{N}=10^{6}$, over a range of the parameter space. The range over which Stirling cycles are swept is outlined (black) along with the critical temperature (red). The outlined region in blue is expanded in b), where a cycle that crosses the transition line is shown. The majority of the expansion stroke occurs in the normal phase and the compression stroke in the condensate phase. The curvature is peaked around $T_c$, giving rise to a boost in the geometric work, which is found by integrating the Berry curvature over the area enclosed by the cycle. c) and d), Work output per particle as a function of $T_l$ and $T_l/T_c$, for a range of system sizes. Each point along the curves correspond to a cycle with a different $T_l$. To emphasise the collective work enhancement we compare this with the same Stirling cycles for a SPE (orange-dashed). The coloured dots in a) and c) indicate the cycles considered in Fig.~\ref{fig: sf result}}
    \label{fig: QS result}
\end{figure}

In the geometric picture, the BEC-induced work surplus appears as a peak in the Berry curvature $\hB{}^{w}$ along the transition line $T=\hbar\omega\sqrt{6\mathcal{N}}/\pi$ in the $\omega-T$ plane, which translates into a peak of the geometric work $\mathcal{W}$ as a function of the base temperature $T_l$ around $T_l\simeq T_c$, see Fig.~\ref{fig: QS result}. To highlight the collective nature of this effect, we have included the geometric work for a SPE,
\begin{equation}
   \mathcal{W}^{1}=2\Bigg(T_h\log\Bigg[\frac{1-\exp\big[\hbar\omega_l/T_h\big]}{1-\exp\big[\hbar\omega_h/T_h\big]}\Bigg]+T_l\log\Bigg[\frac{1-\exp\big[\hbar\omega_h/T_l\big]}{1-\exp\big[\hbar\omega_l/T_l\big]}\Bigg]\Bigg),
    \label{eq: single particle W}
\end{equation}
in the plots 3c) and 3d); for details of the calculation, see \ref{sec: Appendix sho}.

\begin{figure}
    \centering
    \includegraphics[width=\textwidth]{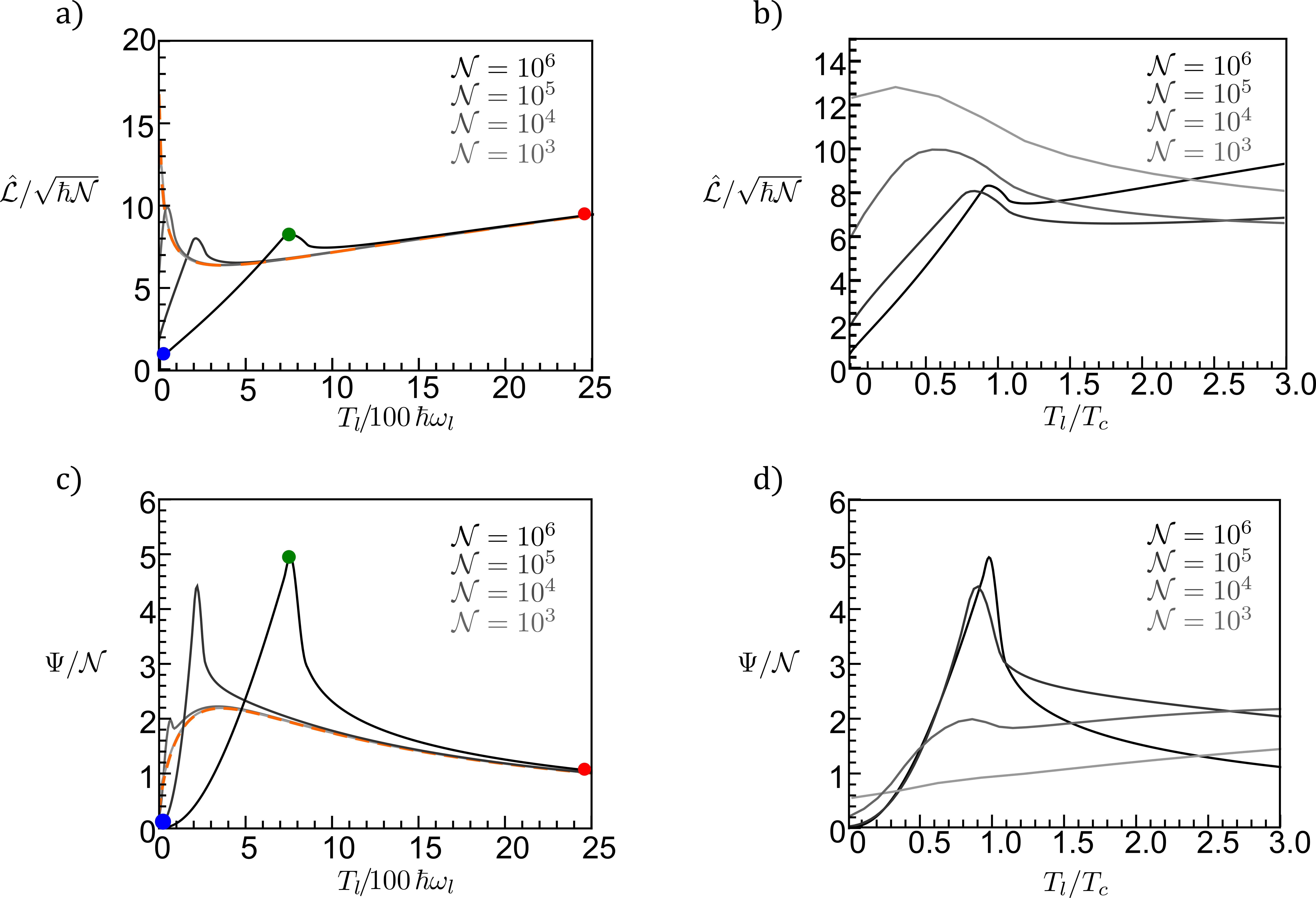}
    \caption{Normalised thermodynamic length and figure of merit $\Psi$ for BEC Stirling cycles. a) and b) show the thermodynamic length for the range of Stirling cycles and system sizes considered in Fig.~\ref{fig: QS result} against $T_l$ and $T_l/T_c$ in comparison with the thermodynamic length of the corresponding single-particle cycles (orange-dashed). Plots c) and d) show the figure $\Psi$ that appears in the trade-off relation \eqref{eq: trade-off}.}
    \label{fig: AR result}
\end{figure}

\begin{figure}
    \centering
    \includegraphics[width=\textwidth]{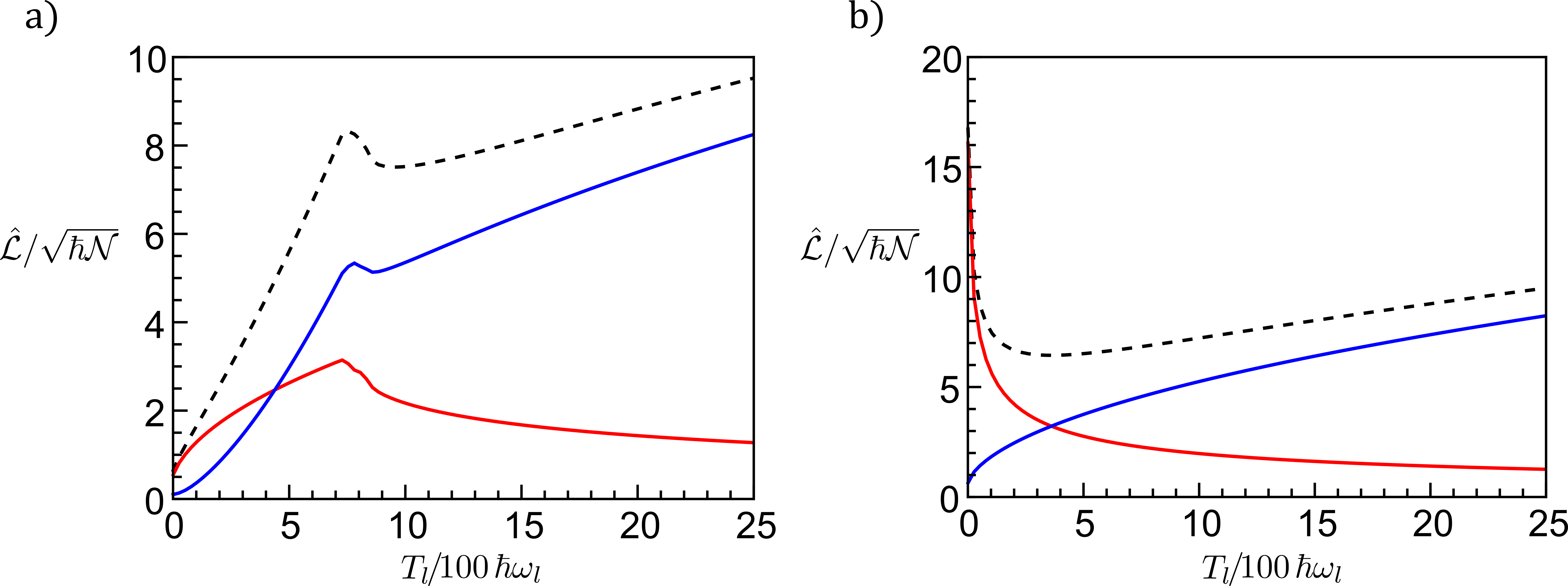}
    \caption{Plots of the normalised thermodynamic length (dashed) and the total contribution arising from isochoric strokes (red) and isothermal strokes (blue) for a) the BEC Stirling engine and b) the SPE.}
    \label{fig: Appendix L}
\end{figure}

\begin{figure}
    \centering
    \includegraphics[width=\textwidth]{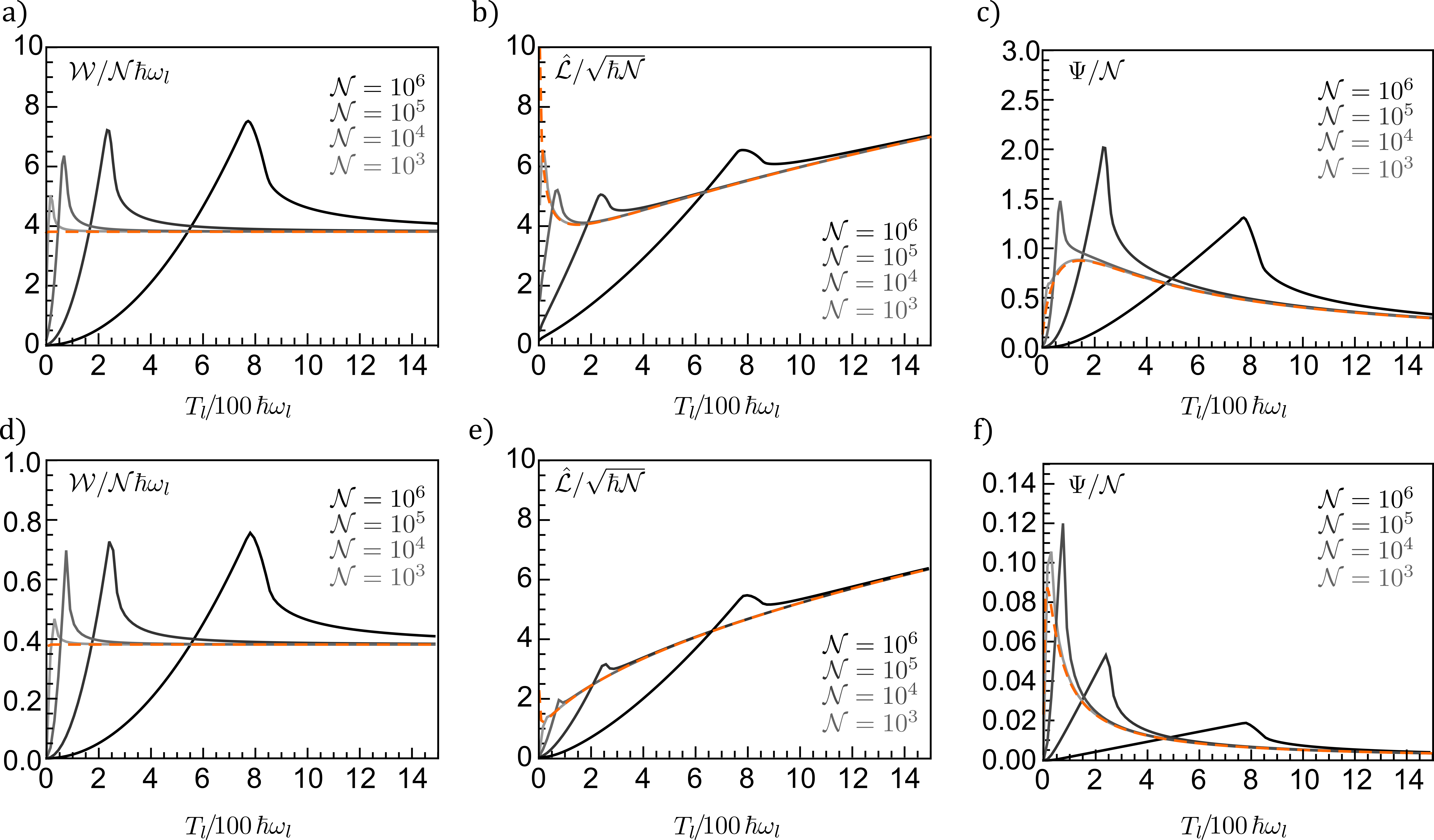}
    \caption{Plots for a) geometric work b) thermodynamic length and c) $\Psi/\mathcal{N}$ for the BEC Stirling engine with cycles where $\Delta T=20\hbar\omega_l$. Plots d)-f) are the same results for cycles with $\Delta T=2\hbar\omega_l$. The results for the SPE (orange-dashed) are shown for comparison.}
    \label{fig: Appendix Delta T=0.2}
\end{figure}

\begin{figure}
    \centering
    \includegraphics[width=0.6\textwidth]{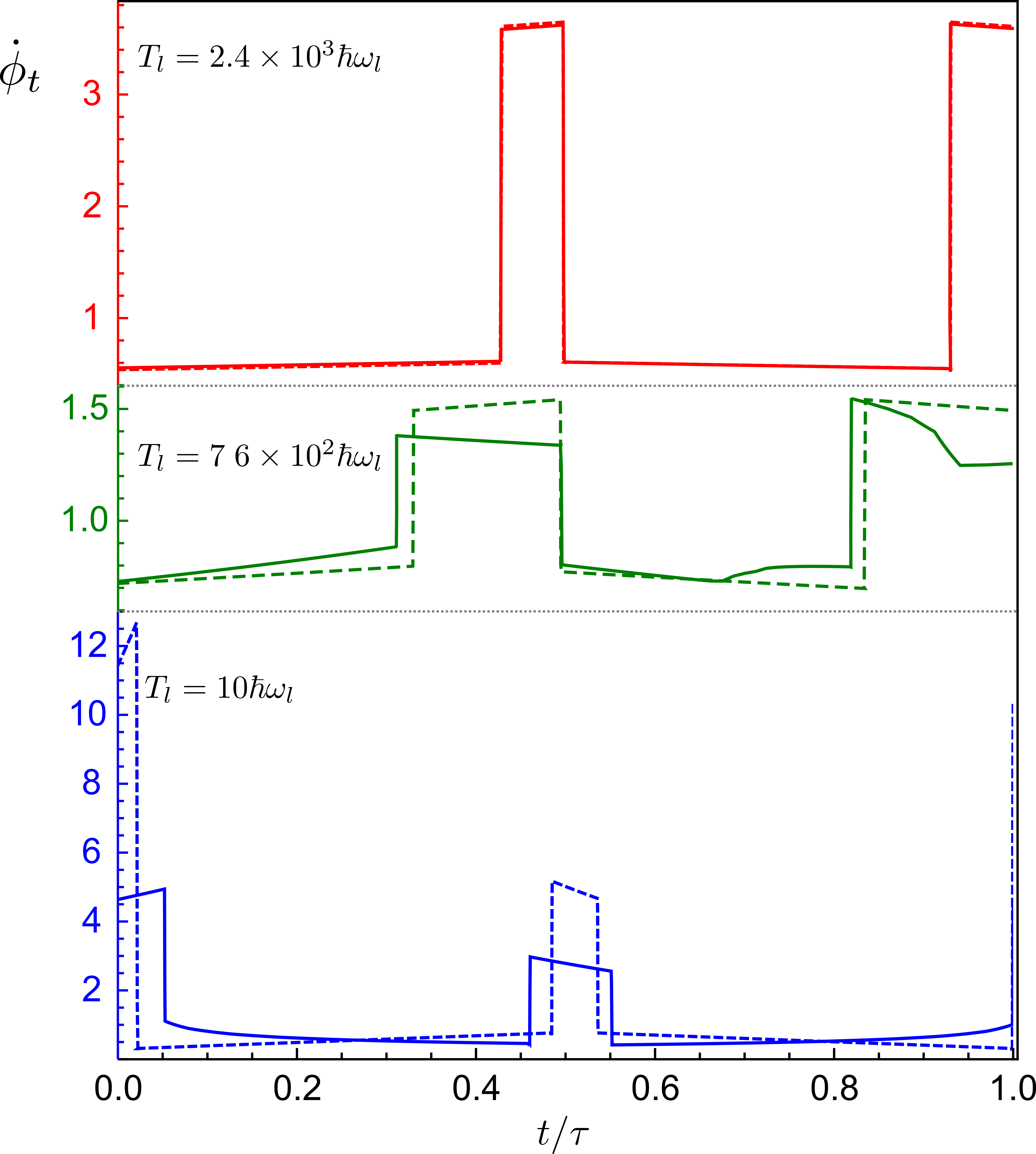}
    \caption{Derivative of the optimal speed functions that minimize the dissipated availability for the SPE (dashed) and BEC engine (solid) with $\mathcal{N}=10^6$. We show results for cycles in the condensate phase (blue), normal phase (red) and across the transition (green), corresponding to the coloured dots in Figs.~\ref{fig: QS result} and \ref{fig: AR result}.}
    \label{fig: sf result}
\end{figure}

For the same cycles, we plot the normalised thermodynamic length $\hat{\mathcal{L}}/\sqrt{\mathcal{N}}$, in Fig.~\ref{fig: AR result}a and b. As with the geometric work per particle, the thermodynamic length has a distinct peak above the work output of SPE in the region where $T_l\simeq T_c$, i.e., when the cycle crosses the phase transition. This result shows that BEC-induced work enhancement comes at the cost of additional dissipation. 
To further understand the behaviour seen in Figs.~\ref{fig: AR result}a and b), we first observe that the thermodynamic length of the single-particle Stirling cycle features a non-monotonic dependence on $T_l$, which arises from the competition between the two contributions, $W$ and $U$, to the dissipated availability, see Eqs.~\eqref{eq: engine W U} and \eqref{eq: engine A}. At high temperatures, the isothermal strokes dominate since compressing the working system at high temperature, and thus at high effective pressure, requires a large work input. At low temperatures, the leading contribution comes from the isochoric strokes, since even small-scale changes in $T$ lead to large variations in the entropy of the system, see Fig.~\ref{fig: Appendix L}. For the many-body cycle, an additional peak in the thermodynamic length emerges, which is caused by the creation and vapourization of a BEC. As the particle number increases, the critical temperature goes up and the peak is offset by the increasing thermodynamic length of the isothermal strokes. If $T_l\ll T_c$ the groundstate is macroscopically occupied throughout the cycle. As a result, both the work exchange during the isothermal strokes and the exchanged thermal energy during the isochoric strokes are suppressed, which leads to a decreasing dissipated availability and thus to a reduced thermodynamic length.
We now turn to the figure of merit $\Psi=(\mathcal{W}/\hat{\mathcal{L}})^2$, which limits the power-efficiency trade-off relation \eqref{eq: trade-off} and is plotted in Figs.~\ref{fig: AR result}c) and d). These plots show a distinct peak in $\Psi/\mathcal{N}$ around the critical temperature, indicating that the BEC-induced work surplus dominates over the corresponding increase in dissipation, in AR. Hence, when operated in the crossover-regime, the many-body Stirling engine can generate significantly more work output per particle at a given efficiency than the corresponding SPE. However, since the peak-value of the thermodynamic length increases faster with $\mathcal{N}$ than the maximum of the geometric work, the peak of $\Psi/\mathcal{N}$ is suppressed at large $\mathcal{N}$. It is therefore possible to determine an optimal system size for a BEC Stirling engine for a specified $\Delta T$. In Fig.~\ref{fig: Appendix Delta T=0.2}, we plot the work per particle, thermodynamic length and figure of merit over the same range of $T_l$ for $\Delta T=20\hbar\omega_l$ and $\Delta T=2\hbar\omega_l$ to illustrate this point.
Finally, the derivatives of the optimal speed functions, which lead to the saturation of the bound \eqref{eq: trade-off}, are shown in Fig.~\ref{fig: sf result}. The plots show $\dot{\phi}_s$ for three cycles, highlighted in Figs.~\ref{fig: QS result} and \ref{fig: AR result}, for the SPE and BEC engines, respectively, with $\mathcal{N}=10^6$. In the high temperature limit, both the SPE and BEC cycle have the same optimal parameterization. In the low temperature limit, the isochoric strokes dominate the dissipated availability and hence are slowed down as to minimize the overall dissipation. Around the critical temperature, the BEC engine exhibits a slight change in the driving rate during the strokes that cross the transition temperature. This feature is not observed for the SPE. 

\FloatBarrier

\section{Concluding Perspectives}
\label{sec: Discussion}
We have presented a geometric framework in which the impact of collective effects on close-to-equilibrium thermodynamic cycles of ideal quantum gases can be quantitatively described. Starting with a grand-canonical picture and invoking the principle equivalence of ensembles, we have derived a set of effective adiabatic-response coefficients, which enable a canonical description of the working system in the language of thermodynamic geometry. This framework includes general expressions for the Berry curvature, which determines the geometric work output of engine cycles, and the thermodynamic length, which determines the minimal dissipation incurred by a given cycle. We have shown that a slowly driven quantum gas engine exhibits an enhanced work output in the quasi-static limit, for cycles that operate across the transition between normal and condensed phase. This result is in agreement with recent work that considers an endoreverisble Otto cycle in a similar setting \cite{Myres2022}. In our approach this enhancement naturally appears as a peak in the corresponding Berry curvature. We have further shown that the thermodynamic length increases in a similar way for cycles that cross the phase boundary, indicating that an additional thermodynamic cost arises from driving the Bose gas across the transition. Nonetheless we find that the power output of the BEC-enhanced engine, which is bounded by a previously derived geometric trade-off relation \cite{Brandner2020}, can be increased at a given efficiency. 

Bose-Einstein condensation is thus an example of a collective quantum effect which can enhance the performance of engine cycles. Further developing our approach could make it possible to explore the role of other many-body effects, such as interactions between particles or quantum phase transitions, in a similar way. The system-independent nature of thermodynamic geometry, which provides, among other results, universal trade-off relations between thermodynamic figures of merit \cite{Brandner2020,Miller2020}, may thus enable a more comprehensive assessment of the role of collective phenomena for the performance of cyclic processes in adiabatic-response.

In this article, we have focused on the geometric trade-off between power and efficiency of ideal quantum-gas engines. As a next step, it would be interesting to include constancy, i.e., inverse power fluctuations, as a third figure of merit, for which similar geometric trade-off relation have been derived \cite{Miller2020}. Whether Bose-Einstein condensation leads to enhanced or diminished work fluctuations in adiabatic-response is generally an open question, which cannot be immediately addressed within our framework for two major reasons. First, quantifying work or power fluctuations would require specifying the protocol for work measurements, which will generally perturb the state of the system. Second, our analysis is based on a grand-canonical treatment which, for sufficiently large systems, is equivalent to a canonical one on the level of mean values of thermodynamic observables. This equivalence does not however necessarily extend to higher-order cumulants. In fact, it is known that the fluctuations of the ground state occupation in an ideal Bose-gas remain macroscopic even at low temperatures in the grand-canonical setting, while they become small in the canonical one \cite{Schmitt2014,Crisanti2019}. One might therefore, expect that our approach would lead to drastically increased work fluctuations below the critical temperature, resulting from macroscopic fluctuations in the groundstate occupation. Overcoming this problem would require formulating a geometric framework for quantum gases in a strictly canonical setting, where the microscopic equations of motion of the working system conserve its particle number. We leave it to future research to find suitable theoretical methods to carry out this program.

On the experimental side, a recently reported realization of an isothermal engine with ultra-cold atoms and tunable interactions that change the quantum statistics of the working system, provides an exciting subject for future studies \cite{Koch2022}. Here, it is a priori unclear how a geometric framework should be constructed, since, at least in an effectively non-interacting model, the statistics of quantum particles cannot be continuously tuned between Bose-Einstein and Fermi-Dirac. Describing this process on a microscopic level, would require interactions between particles to be taken into account, to enable the transition from a molecular BEC to a Fermi sea. In a more general perspective, extending our dynamical approach to more realistic interacting systems would be a significant development in its own right, which would give access to a new class of collective effects that cannot be observed in ideal quantum gases.

\ack
KB acknowledges support from the University of Nottingham through a Nottingham Research Fellowship. This work was supported by the Medical Research Council [grant number MR/S034714/1]; and the Engineering and Physical Sciences Research Council [grant number EP/V031201/1]. KS was supported by Grants-in-Aid for Scientific Research (JP19H05603 and JP19H05791). TP was supported by the Academy of Finland (grant number 312057) and the Nokia Industrial Doctoral School in Quantum Technology

\appendix
\section{Single harmonic oscillator engine}
\newcommand{\at}{\boldsymbol{\mathrm{a}}_{\lmd}}
\newcommand{\adt}{\boldsymbol{\mathrm{a}}^{\dagger}_{\lmd}}
\newcommand{\rla}{\langle\adt\at\rangle}
In this appendix, we derive the expression for the quasi-static work output and thermodynamic length for a single-particle working system with mass $m$ in a 2d-harmonic isotropic trap. 
\subsection{Quasi-static work output}
\label{sec: Appendix sho}
The Hamiltonian for this system reads
\begin{equation}
    H^{1}_{\omega}=\hbar\omega\big(\ad\adn+1\big),
\end{equation}
with the creation and annihilation operators taking the form $\adn\equiv\sqrt{m\omega/2}(\boldsymbol{\mathrm{x}}+i\boldsymbol{\mathrm{p}}/m\omega)$, $\ad\equiv\sqrt{m\omega/2}(\boldsymbol{\mathrm{x}}-i\boldsymbol{\mathrm{p}}/m\omega)$, where $\boldsymbol{\mathrm{x}}$ and $\boldsymbol{\mathrm{p}}$ are the 2-dimensional position and momentum operators, and $\omega$ is the trap strength. The free energy is
\begin{align}
    \mathcal{F}_{\lmd}^{1}=-T\log{\text{tr}\big[\exp{[-H^1_\omega/T]}\big]}&=-T\log{\bigg[\sum_{n=0}^{\infty}(n+1)\exp{\big[-\hbar\omega(n+1)/T\big]}\bigg]}\nonumber\\
    &=-\hbar\omega+2T\log{\big[\exp\big[\hbar\omega/T\big]-1\big]}.
\end{align}

The work output for any engine cycle where the trap strength $\omega$ and temperature $T$ are varied periodically is given by Eq.~\eqref{eq: engine W U}. In the quasi-static limit, the thermodynamic force is given by $f^w_t\rightarrow\mathcal{F}^{w}_{\lmd_t}=-\partial_w \mathcal{F}_{\lmd}$ and thus the quasi-static work becomes
\begin{equation}
   \mathcal{W}^{1}=2\Bigg(T_h\log\Bigg[\frac{1-\exp\big[\hbar\omega_l/T_h\big]}{1-\exp\big[\hbar\omega_h/T_h\big]}\Bigg]+T_l\log\Bigg[\frac{1-\exp\big[\hbar\omega_h/T_l\big]}{1-\exp\big[\hbar\omega_l/T_l\big]}\Bigg]\Bigg).
\end{equation}

\subsection{Thermodynamic length}
To calculate the thermodynamic length, we require the AR coefficients, which have to be derived from the equilibrium dynamics of the system. To this end, we introduce the Lindblad master equation
\begin{equation}
    \dot{\rho}_t=\boldsymbol{\mathrm{L}}^1_t\rho_t \quad\text{with}\quad\boldsymbol{\mathrm{L}}^1_t=-\frac{i}{\hbar}\big[H^1_{\omega_t},\rho_t\big] + \boldsymbol{\mathrm{D}}^{1}_{\Lmd_t}, 
    \label{eq: Lin ME sho}
\end{equation}
which describes a single harmonic oscillator, coupled to a thermal bath at temperature $T$. The unitary evolution is described by the first term in \eqref{eq: Lin ME sho} and the dissipation super-operator is given by 
\begin{equation}
    \boldsymbol{\mathrm{D}}^{1}_{\Lmd}\boldsymbol{\cdot}\equiv\kappa\bigg[\mathfrak{n}_{\bu{n},\lmd}\bigg(\at\boldsymbol{\cdot}\adt-\frac{1}{2}\big\{\boldsymbol{\cdot},\adt\at\big\}\bigg)+\big(\mathfrak{n}_{\bu{n},\lmd}+1\big)\bigg(\adt\boldsymbol{\cdot}\at-\frac{1}{2}\big\{\boldsymbol{\cdot},\at\adt\big\}\bigg)\bigg]
\end{equation}
with the adjoint
\begin{equation}
    \boldsymbol{\mathrm{D}}_{\lmd}^{1\dagger}\boldsymbol{\cdot}\equiv\kappa\bigg[\big(\mathfrak{n}_{\bu{n},\lmd}+1\big)\bigg(\adt\boldsymbol{\cdot}\at-\frac{1}{2}\big\{\boldsymbol{\cdot},\adt\at\big\}\bigg)+\mathfrak{n}_{\bu{n},\lmd}\bigg(\at\boldsymbol{\cdot}\adt-\frac{1}{2}\big\{\boldsymbol{\cdot},\at\adt\big\}\bigg)\bigg].
\end{equation}
Here, the rate $\kappa>0$ sets the relaxation timescale. The dissipation super-operator thus obeys the detailed balance condition
\begin{equation}
    \boldsymbol{\mathrm{D}}^{1}_{\lmd}\varrho_{\Lmd}= \varrho_{\lmd}\boldsymbol{\mathrm{D}}_{\Lmd}^{1\dagger} \quad \text{with}\quad 
    \varrho_{\lmd}=\exp\big[-\big(H^{1}_{\omega}-\mathcal{F}^1_{\lmd}\big)/T\big]
\end{equation}
being the instantaneous stationary solution of Eq.~\eqref{eq: Lin ME sho}.

We now make the ansatz
\begin{equation}
    \rho_t\simeq\varrho_{\lmd_t}+\sum_{a}\varrho^{a}_{\lmd_t}\dot{\lambda}_t^a,
    \label{eq: state ansatz}
\end{equation}
for the periodic state, which we substitute into Eq.~\eqref{eq: Lin ME sho}. Expanding the left hand side in time derivatives of the control parameters and equating coefficients of $\dot{\lambda}^a$ yields
\begin{equation}
    \varrho_{\lmd_t}^a=\big(\boldsymbol{\mathrm{L}}^1_t\big)^{-1}\partial_{a}\varrho_{\lmd_t},
    \label{eq: inv L}
\end{equation}
where $\big(\boldsymbol{\mathrm{L}}^1_t\big)^{-1}$ is the psuedo-inverse of the Lindblad generator $\boldsymbol{\mathrm{L}}^1_t$. From here we drop the time arguments for notational convenience. The derivatives of the Gibbs state with respect to the control parameters are
\begin{align}
    \partial_w\varrho_{\lmd}&=-\varrho_{\lmd}\bigg[\frac{\hbar}{T}(\adt\at-\rla)+\frac{1}{4\omega}\sum_{\nu=x,y}\big(\mathrm{a}_{\nu,\lmd}^{\dagger2}\big(1-\exp[2\hbar\omega/T]\big)\nonumber\\
    &\qquad\qquad\qquad\qquad\qquad\qquad\qquad\qquad-\mathrm{a}_{\nu,\lmd}^2\big(\exp[-2\hbar\omega/T]-1\big)\big)\bigg],\\
    \partial_u\varrho_{\lmd}&=\frac{\hbar\omega}{T^2}\varrho_{\lmd}\Big(\adt\at-\rla\Big),
\end{align}
where angular brackets denote the thermal average, i.e., $\langle X\rangle=\text{tr}[X\varrho_{\lmd}]$, which we substitute into Eq.~\eqref{eq: inv L} to obtain
\begin{align}
    \varrho^w_{\lmd}&=-\frac{\hbar}{\kappa T}\big(\adt\at-\rla\big)\nonumber\\
    &\qquad+\frac{1}{4\omega}\sum_{\nu=x,y}\Bigg[\frac{\big(\exp{[-2\hbar\omega/T]}-1\big)}{\kappa+2i\omega}\mathrm{a}_{\nu,\lmd}^{\dagger 2}-\frac{\big(1-\exp{[2\hbar\omega/T]}\big)}{\kappa-2i\omega}\mathrm{a}_{\nu,\lmd}^2\Bigg]\varrho_{\lmd},\label{eq: sho ansatz1}\\
    \varrho^u_{\lmd}&=\frac{\hbar\omega}{\kappa T^2}\big(\adt\at-\rla\big)\varrho_{\lmd}.
    \label{eq: sho ansatz2}
\end{align}

In AR, the thermodynamic forces admit the expansion up to first order in the driving rates, 
\begin{equation}
    f^{a}_t\simeq\mathcal{F}^{a}_{\lmd}+\sum_{b}R^{ab}_{\lmd}\dot{\lambda}^{b},
    \label{eq: gen f sho}
\end{equation}
where $R^{ab}_{\lmd}$ are the AR coefficients that appear in the expression for the thermodynamic length. Upon substituting the state ansatz \eqref{eq: state ansatz} into \eqref{eq: gen f sho} and equating coefficients of $\dot{\lambda}^b$, it is straightforward to show that the AR coefficients are given by
\begin{equation}
    R^{wb}_{\lmd}=-\text{tr}\big[(\partial_w H^1_\omega)\varrho^b_{\lmd}\big]\quad\text{and}\quad R^{ub}=-\text{tr}\big[\varrho^b_{\lmd}\log{[\varrho_{\lmd}]}\big].
    \label{eq: therm f exp sho}
\end{equation}
We can therefore determine the coefficients $R^{ab}_{\lmd}$ by inserting the results from Eqs.~\eqref{eq: sho ansatz1} and \eqref{eq: sho ansatz2} into Eq.~\eqref{eq: therm f exp sho}, which yields
\begin{align}
    &R^{ww}_{\lmd}=\frac{\hbar^2}{\kappa T}\llangle \adt\at \rrangle-\frac{\hbar\big(\exp[2\hbar\omega/T]-1\big)}{4\omega (\kappa^2+4\omega^2)}\sum_{\nu=x,y}\langle\mathrm{a}_{\nu,\lmd}^{\dagger 2}\mathrm{a}_{\nu,\lmd}^2\rangle\\
    &R^{uu}_{\lmd}=\frac{\hbar^2\omega^2}{\kappa T^3}\llangle \adt\at \rrangle,\\
    &R^{wu}_{\lmd}=R^{uw}_{\lmd}=\frac{\hbar^2\omega}{\kappa T^2}\llangle \adt\at \rrangle
\end{align}
with the notation $\llangle X \rrangle=\langle X\rangle^2-\langle X^2\rangle$. Taking the expectation values of the creation and annihilation operators and their squares, we find the exact expressions for the AR coefficients,
\begin{align}
    &R^{ww}_{\lmd}=\frac{\hbar^2}{\kappa T}\frac{2\exp{[\hbar\omega/T]}}{\big(\exp{[\hbar\omega/T]}-1\big)^2}\Bigg[1-\frac{T\big(\exp[\hbar\omega/T]-\exp[-\hbar\omega/T]\big)}{2\hbar\omega (1+(2\omega/\kappa)^2)}\Bigg]\\
    &R^{uu}_{\lmd}=-\frac{\hbar^2\omega^2}{\kappa T^3}\frac{2\exp{[\hbar\omega/T]}}{\big(\exp{[\hbar\omega/T]}-1\big)^2},\\
    &R^{wu}_{\lmd}=R^{uw}_{\lmd}=-\frac{\hbar^2\omega}{\kappa T^2}\frac{2\exp{[\hbar\omega/T]}}{\big(\exp{[\hbar\omega/T]}-1\big)^2}.
\end{align}
The thermodynamic length follows by inserting these results into the formula
\begin{equation}
    \mathcal{L}=\oint_\gamma \ \Bigg[\sum_{a,b}g^{ab}_{\lmd}d\lambda^a d\lambda^b\Bigg]^{1/2},\quad \text{where} \quad g^{ab}_{\lmd}=-(R^{ab}_{\lmd}+R^{ba}_{\lmd})/2,
    \label{eq: L sho}
\end{equation}
upon specifying the path $\gamma$.

\section{Reduction coefficients for BEC engine}
\label{sec: Appendix B}
In this appendix, we outline the analytic expressions for the coefficients $M^{\alpha\beta}_{\Lmd}$ which appear in Eq.~\eqref{eq: Reduced R} for the harmonically trapped Bose gas. Given the expressions for the grand potential, $\Phi_{\Lmd}$, and the three thermodynamic forces, $\mathcal{F}^{\alpha}_{\Lmd}$,
\begin{align}
    \Phi_{\Lmd}=&-T\Bigg[\Li{1}{z}+\frac{2\Li{2}{zq}-\Li{2}{zq^{2}}}{(\hbar\omega/T)}+\frac{2\Li{3}{zq}-\Li{3}{zq^{2}}}{(\hbar\omega/T)^2}\Bigg],\\
    \mathcal{F}_{\Lmd}^w=&-\hbar\Bigg[\Li{0}{z}+\frac{4\Li{1}{zq}-3\Li{1}{zq^2}}{\big(\hbar\omega/T\big)}+\frac{6\Li{2}{zq}-4\Li{2}{zq^2}}{\big(\hbar\omega/T\big)^2}\nonumber\\
    &\qquad+\frac{4\Li{2}{zq}-2\Li{2}{zq^2}}{\big(\hbar\omega/T\big)^3}\Bigg],\\
    \mathcal{F}_{\Lmd}^u=\ &\frac{1}{T}\Bigg[\Phi_{\Lmd}+\xi_{1}\Li{0}{z}+\frac{\xi_22\Li{1}{zq}-\xi_3\Li{1}{zq^2}}{\big(\hbar\omega/T\big)}+\frac{2\xi_2\Li{2}{zq}-\xi_3\Li{2}{zq^2}}{\big(\hbar\omega/T\big)^2}\Bigg]\nonumber\\
    &+\frac{2\Li{2}{zq}-\Li{2}{zq^2}}{\big(\hbar\omega/T\big)}+\frac{4\Li{2}{zq}-2\Li{2}{zq^2}}{\big(\hbar\omega/T\big)^2},\\
    \mathcal{F}_{\Lmd}^z=&\ \Li{0}{z}+\frac{2\Li{1}{zq}-\Li{1}{zq^{2}}}{(\hbar\omega/T)}+\frac{2\Li{2}{zq}-\Li{2}{zq^{2}}}{(\hbar\omega/T)^2},
\end{align}
we find the coefficients $M^{\alpha\beta}_{\Lmd}=\partial_{\alpha}\mathcal{F}^{\beta}_{\Lmd}=-\partial_{\alpha}\partial_{\beta}\Phi_{\Lmd}$,
\begin{align}
    M^{ww}_{\Lmd}&=\frac{\hbar^2}{T}\Bigg[\Li{-1}{z}+\frac{8\Li{0}{zq}-9\Li{0}{zq^2}}{\big(\hbar\omega/T\big)}+\frac{16\Li{1}{zq}-15\Li{1}{zq^2}}{\big(\hbar\omega/T\big)^2}\nonumber\\
    &\qquad\qquad+\frac{20\Li{2}{zq}-14\Li{2}{zq^2}}{\big(\hbar\omega/T\big)^3}+\frac{12\Li{2}{zq}-6\Li{2}{zq^2}}{\big(\hbar\omega/T\big)^4}\Bigg],\\
    M^{uu}_{\Lmd}&=-\frac{2}{T}\mathcal{F}^{u}_{\Lmd}+\frac{1}{T^3}\Bigg[\xi_1^2\Li{-1}{z}+\frac{2\xi_{2}^2\Li{0}{zq}-\xi^2_3\Li{0}{zq^2}}{\big(\hbar\omega/T\big)}+\frac{2\xi_{2}^2\Li{1}{zq}-\xi^2_3\Li{1}{zq^2}}{\big(\hbar\omega/T\big)^2}\Bigg]\nonumber\\
    &\qquad\qquad+\frac{2}{T^2}\Bigg[\frac{2\xi_{2}\Li{1}{zq}-\xi_3\Li{1}{zq^2}}{\big(\hbar\omega/T\big)}+\frac{4\xi_{2}\Li{2}{zq}-2\xi_3\Li{2}{zq^2}}{\big(\hbar\omega/T\big)^2}\Bigg]\nonumber\\
    &\qquad\qquad+\frac{1}{T}\Bigg[\frac{2\Li{2}{zq}-\xi_3\Li{2}{zq^2}}{\big(\hbar\omega/T\big)}+\frac{2\xi_{2}\Li{2}{zq}-\xi_3\Li{2}{zq^2}}{\big(\hbar\omega/T\big)^2}\Bigg],\\
     M^{zz}_{\Lmd}&=\frac{1}{T}\Bigg[\Li{-1}{z}+\frac{2\Li{0}{z}-\Li{0}{zq^2}}{\big(\hbar\omega/T\big)}+\frac{2\Li{1}{z}-\Li{1}{zq^2}}{\big(\hbar\omega/T\big)^2}\Bigg],\\
    M^{uw}_{\Lmd}&=M^{wu}_{\Lmd}=-\frac{1}{T}\mathcal{F}_{\Lmd}^w-\frac{\hbar}{T}\Bigg[\Li{0}{z}+\frac{6\Li{2}{zq}-4\Li{2}{zq^2}}{\big(\hbar\omega/T\big)^2}+\frac{8\Li{3}{zq}-4\Li{3}{zq^2}}{\big(\hbar\omega/T\big)^3}\Bigg]\nonumber\\
    &\qquad\qquad-\frac{\hbar}{T^2}\Bigg[\xi_1\Li{-1}{z}+\frac{4\xi_2\Li{0}{zq}-3\xi_3\Li{0}{zq^2}}{\big(\hbar\omega/T\big)}+\frac{6\xi_2\Li{1}{zq}-4\xi_3\Li{1}{zq^2}}{\big(\hbar\omega/T\big)^2}\nonumber\\
    &\qquad\qquad\qquad\qquad+\frac{4\xi_2\Li{2}{zq}-2\xi_3\Li{2}{zq^2}}{\big(\hbar\omega/T\big)^3}\Bigg],\\
    M^{zw}_{\Lmd}&=M^{wz}_{\Lmd}=-\frac{\hbar}{T}\Bigg[\Li{-1}{z}+\frac{4\Li{0}{zq}-2\Li{0}{zq^2}}{\big(\hbar\omega/T\big)}+\frac{6\Li{1}{zq}-4\Li{1}{zq^2}}{\big(\hbar\omega/T\big)^2}\nonumber\\
    &\qquad\qquad\qquad\qquad+\frac{4\Li{2}{zq}-2\Li{2}{zq^2}}{\big(\hbar\omega/T\big)^3}\Bigg],\\
    M^{zu}_{\Lmd}&=M^{uz}_{\Lmd}=-\frac{1}{T}\mathcal{F}^{z}_{\Lmd}-\frac{1}{T}\Bigg[\Li{0}{z}-\frac{2\Li{2}{zq}-\Li{2}{zq^2}}{\big(\hbar\omega/T\big)^2}\Bigg]\nonumber\\
    &\qquad\qquad+\frac{1}{T^2}\Bigg[\xi_{1}\Li{-1}{z}+\frac{2\xi_{2}\Li{0}{z}-\xi_{3}\Li{0}{zq^2}}{\big(\hbar\omega/T\big)}+\frac{2\xi_{2}\Li{1}{z}-\xi_{3}\Li{1}{zq^2}}{\big(\hbar\omega/T\big)^2}\Bigg].
    \label{eq: Mexpanded}
\end{align}
Here, we recall the definition $\xi_n=\hbar\omega n - \mu$. The coefficients $M^{\alpha\beta\gamma}_{\Lmd}=\partial_{\alpha}M^{\beta\gamma}_{\Lmd}$ can be obtained by taking further derivatives. 

\section{Primary AR coefficients for BEC engine}
\label{sec: Appendix D}
\newcommand{\Lin}{\mathrm{Xi}_{1}(zq)}
\newcommand{\Linn}{\mathrm{Xi}_{2}(zq)}
\newcommand{\Linnn}{\mathrm{Xi}_{3}(zq)}
\newcommand{\Bi}[1]{\mathrm{Yi}(#1 q)}
In this appendix, we outline the approximate expressions for the primary AR coefficients of the BEC engine. To calculate the coefficients explicitly, we follow the same approach as for the grand-potential, $\Phi_{\Lmd}$, in Sec.~\ref{sec: equilibrium BEC}. 

The primary AR coefficients are given by Eq.~\eqref{eq: AR coeff BEC}, where
\begin{equation}
   \mathfrak{n}_{n,\Lmd}\big(\mathfrak{n}_{n,\Lmd}+1\big)=\frac{zq^n}{(1-zq^n)^2}\quad\text{and}\quad\mathfrak{n}_{n+2,\Lmd}\big(\mathfrak{n}_{n,\Lmd}+1\big)=\bigg[\frac{zq^n}{1-zq^n}-\frac{z'q^n}{1-z'q^n}\bigg]
    \label{eq: two mathfraks}
\end{equation}
and $z'=\exp{\big[(\mu-3\hbar\omega)/T\big]}$. The coefficients $g_{n}$ and $D_{n,\Lmd}$, in Eq.~\eqref{eq: AR coeff BEC}, have the form $(n+1)^p$ with $p\in \{0,1,2,3\}$. We therefore introduce the general expression
\begin{equation}
    \sum_{n=0}^{\infty}(n+1)^p\frac{zq^n}{(1-zq^n)^2}=\mathrm{Xi}_{p}(zq)+\frac{z}{(1-z)^2}
    \label{eq: general polynomial}
\end{equation}
where,
\begin{equation}
    \mathrm{Xi}_{p}(zq)=\sum_{m=0}^{\infty}(m+2)^p\frac{zq^{(m+1)}}{\big(1-zq^{(m+1)}\big)^2}.
\end{equation}
Using that $\sum_{x=0}^{\infty}\big(1-x\big)^{-2}=\sum_{j=0}^{\infty}c_j x^j $ for $x<1$, where $c_j=j+1$, we find
\begin{align}
     \mathrm{Xi}_{p}(zq)=\sum_{j=0}^{\infty}c_j\big(zq\big)^{j+1}&\sum_{m=0}^{\infty}\big(m+2\big)^p q^{(j+1)m}\nonumber\\
    =\sum_{j=0}^{\infty}c_j\big(zq\big)^{j+1}&\bigg[(X_0^p+X_1^p+X_2^p+X_3^p)+\frac{(8X_3^p+4X_2^p+2X_1^p+X_0^p)}{q^{j+1}-1}\nonumber\\
    & \qquad+\frac{(19X_p^3+5X_2^p+X_0^p)}{\big(q^{j+1}-1\big)^2}+\frac{2(9X_3^p+X_2^p)}{\big(q^{j+1}-1\big)^{3}}+\frac{6X_3^p}{\big(q^{j+1}-1\big)^4}\bigg]
    \label{eq: Ai}
\end{align}
where $X_n^p=1^{p-n}\binom{p}{n}$. 

So far no approximations have been made. We now expand the denominators in Eq.~\eqref{eq: Ai} to leading order in $T/\hbar\omega$, which is justified since $\hbar\omega\ll T$. We thus arrive at the approximate expressions
\begin{align}
    \mathrm{Xi}_{p}(zq)&\simeq (X_0^p+X_1^p+X_2^p+X_3^p)\Li{-1}{zq}+\frac{(X_0^p+X_1^p+X_2^p+X_3^p)}{\big(\hbar\omega/T\big)}\Li{0}{zq}\nonumber\\
    &\quad+\frac{(3X_3^p+2X_2^p+X_1^p)}{\big(\hbar\omega/T\big)^2}\Li{1}{zq}+\frac{(6X_3^p+2X_2^p)}{\big(\hbar\omega/T\big)^3}\Li{2}{zq}+\frac{6X_3^p}{\big(\hbar\omega/T\big)^4}\Li{3}{zq}.
\end{align}

With this result, all coefficients except $R^{ww}_{\Lmd}$ can be fully expressed in terms of sums of ploylogarithms. To calculate $R^{ww}_{\Lmd}$, we must determine
\begin{align}
    \sum_{n=0}^{\infty}h_n\mathfrak{n}_{n+2,\Lmd}\big(\mathfrak{n}_{n,\Lmd}+1\big)=4\bigg[\frac{z}{1-z}-\frac{z'}{1-z'}\bigg]+
    \mathrm{Yi}(zq)-\mathrm{Yi}(z'q),
\end{align}
where
\begin{equation}
    \mathrm{Yi}(zq)=\frac{2}{3}\sum_{m=0}^{\infty}(m+2)(m+3)(m+4)\bigg[\frac{zq^{(m+1)}}{1-zq^{(m+1)}}\bigg].
\end{equation}
Following the same approach as for the previous terms, we find
\begin{align}
     \mathrm{Yi}(zq)\simeq &\ \frac{2}{3}\bigg[24\Li{0}{zq}+\frac{24}{\big(\hbar\omega/T\big)}\Li{1}{zq}+\frac{26}{\big(\hbar\omega/T\big)^2}\Li{2}{zq}+\frac{18}{\big(\hbar\omega/T\big)^3}\Li{3}{zq}\nonumber\\
    &\quad+\frac{6}{\big(\hbar\omega/T\big)^4}\Li{4}{zq}\bigg].
\end{align}
With these general expressions it is now straightforward to write the approximate AR coefficients in the form
\begin{align}
    R^{ww}_{\Lmd}&=-\frac{\hbar^2}{\kappa T}\bigg[\Li{-1}{zq}+\Linn+\frac{T}{\hbar\omega\big(1+(2\omega/\kappa)^2\big)}\Big(\Li{0}{zq}-\Li{0}{z'q}\nonumber\\
    &\qquad\qquad+\Bi{z}-\Bi{z'}\Big)\bigg],\\
    R^{uu}_{\Lmd}&=-\frac{1}{\kappa T^3}\bigg[\xi_{1}^2\Li{-1}{zq}+\big(\hbar\omega\big)^2\Linnn-2\hbar\omega\mu\Linn+\mu^2\Lin\bigg],\\
   R^{zz}_{\Lmd}&=-\frac{1}{\kappa T}\bigg[\Li{-1}{zq}+\Lin\bigg],\\
   R^{uw}_{\Lmd}&=R^{wu}_{\Lmd}=\frac{\hbar}{\kappa T^2}\bigg[\xi_{1}\Li{-1}{zq}+\hbar\omega\Linnn-\mu\Linn\bigg],\\
   R^{zw}_{\Lmd}&=R^{wz}_{\Lmd}=\frac{\hbar}{\kappa T}\bigg[\Li{-1}{zq}+\Linn\bigg],\\
   R^{zu}_{\Lmd}&=R^{uz}_{\Lmd}=-\frac{1}{\kappa T^2}\bigg[\xi_1 \Li{-1}{zq}+\hbar\omega\Linn-\mu\Lin\bigg].
\end{align}
Together with Eqs.~\eqref{eq: Mexpanded}, these results can be substituted into Eq.~\eqref{eq: Reduced R} to obtain the reduced AR coefficients.

\bibliographystyle{iopart-num}
\bibliography{library}

\providecommand{\newblock}{}
\begin{thebibliography}{10}
\expandafter\ifx\csname url\endcsname\relax
  \def\url#1{{\tt #1}}\fi
\expandafter\ifx\csname urlprefix\endcsname\relax\def\urlprefix{URL }\fi
\providecommand{\eprint}[2][]{\url{#2}}

\bibitem{Seifert2012}
Seifert U 2012 {\em Rep. Prog. Phys.\/} {\bf 75} 126001
  \urlprefix\url{https://dx.doi.org/10.1088/0034-4885/75/12/126001}

\bibitem{Vinjanampathy2016}
Vinjanampathy S and Anders J 2016 {\em Contemp. Phys.\/} {\bf 57} 545
  \urlprefix\url{https://doi.org/10.1080/00107514.2016.1201896}

\bibitem{Sivak2012}
Sivak D~A and Crooks G~E 2012 {\em Phys. Rev. Lett.\/} {\bf 108} 190602
  \urlprefix\url{https://link.aps.org/doi/10.1103/PhysRevLett.108.190602}

\bibitem{Crooks2012}
Zulkowski P~R, Sivak D~A, Crooks G~E and DeWeese M~R 2012 {\em Phys. Rev. E\/}
  {\bf 86} 041148
  \urlprefix\url{https://link.aps.org/doi/10.1103/PhysRevE.86.041148}

\bibitem{Rotskoff2015}
Rotskoff G~M and Crooks G~E 2015 {\em Phys. Rev. E\/} {\bf 92} 060102
  \urlprefix\url{https://link.aps.org/doi/10.1103/PhysRevE.92.060102}

\bibitem{Zulkowski2015}
Zulkowski P~R and DeWeese M~R 2015 {\em Phys. Rev. E\/} {\bf 92} 032117
  \urlprefix\url{https://link.aps.org/doi/10.1103/PhysRevE.92.032117}

\bibitem{Zulkowski2015b}
Zulkowski P~R and DeWeese M~R 2015 {\em Phys. Rev. E\/} {\bf 92} 032113
  \urlprefix\url{https://link.aps.org/doi/10.1103/PhysRevE.92.032113}

\bibitem{Cavina2017}
Cavina V, Mari A and Giovannetti V 2017 {\em Phys. Rev. Lett.\/} {\bf 119}
  050601
  \urlprefix\url{https://link.aps.org/doi/10.1103/PhysRevLett.119.050601}

\bibitem{Miller2019}
Miller H~J~D, Scandi M, Anders J and Perarnau-Llobet M 2019 {\em Phys. Rev.
  Lett.\/} {\bf 123} 230603
  \urlprefix\url{https://link.aps.org/doi/10.1103/PhysRevLett.123.230603}

\bibitem{Abiuso2020}
Abiuso P, Miller H~J~D, Perarnau-Llobet M and Scandi M 2020 {\em Entropy\/}
  {\bf 22} \urlprefix\url{https://www.mdpi.com/1099-4300/22/10/1076}

\bibitem{Blaber2020}
Blaber S and Sivak D~A 2020 {\em J. Chem. Phys.\/} {\bf 153} 244119
  \urlprefix\url{https://doi.org/10.1063/5.0033405}

\bibitem{blaber2022optimal}
Blaber S and Sivak D~A 2022 Optimal control in stochastic thermodynamics
  \urlprefix\url{https://arxiv.org/abs/2212.00706}

\bibitem{Mehboudi2022}
Mehboudi M and Miller H~J~D 2022 {\em Phys. Rev. A\/} {\bf 105} 062434
  \urlprefix\url{https://link.aps.org/doi/10.1103/PhysRevA.105.062434}

\bibitem{Abiuso2022}
Abiuso P, Holubec V, Anders J, Ye Z, Cerisola F and Perarnau-Llobet M 2022 {\em
  J. Phys. Commun.\/} {\bf 6} 063001
  \urlprefix\url{https://doi.org/10.1088/2399-6528/ac72f8}

\bibitem{Scandi2022}
Scandi M, Barker D, Lehmann S, Dick K~A, Maisi V~F and Perarnau-Llobet M 2022
  Constant dissipation rate is optimal for thermodynamic protocols:
  experimental implementation of landauer erasure through thermodynamic length
  \urlprefix\url{https://arxiv.org/abs/2209.01852}

\bibitem{Weinberg2017}
Weinberg P, Bukov M, D’Alessio L, Polkovnikov A, Vajna S and Kolodrubetz M
  2017 {\em Phys. Rep.\/} {\bf 688} 1--35
  \urlprefix\url{https://www.sciencedirect.com/science/article/pii/S0370157317301412}

\bibitem{Weinhold1975}
Weinhold F 1975 {\em J. Chem. Phys.\/} {\bf 63} 2479--2483
  \urlprefix\url{https://doi.org/10.1063/1.431689}

\bibitem{Andresen1984}
Andresen B, Salamon P and Berry R~S 1984 {\em Physics Today\/} {\bf 37} 62--70
  \urlprefix\url{https://doi.org/10.1063/1.2916405}

\bibitem{Brody1995}
Brody D and Rivier N 1995 {\em Phys. Rev. E\/} {\bf 51} 1006--1011
  \urlprefix\url{https://link.aps.org/doi/10.1103/PhysRevE.51.1006}

\bibitem{Ruppeiner1995}
Ruppeiner G 1995 {\em Rev. Mod. Phys.\/} {\bf 67} 605--659
  \urlprefix\url{https://link.aps.org/doi/10.1103/RevModPhys.67.605}

\bibitem{Salmon1983}
Salamon P and Berry R~S 1983 {\em Phys. Rev. Lett.\/} {\bf 51}
  \urlprefix\url{https://link.aps.org/doi/10.1103/PhysRevLett.51.1127}

\bibitem{Crooks2007}
Crooks G~E 2007 {\em Phys. Rev. Lett.\/} {\bf 99} 100602
  \urlprefix\url{https://link.aps.org/doi/10.1103/PhysRevLett.99.100602}

\bibitem{Machta2015}
Machta B~B 2015 {\em Phys. Rev. Lett.\/} {\bf 115} 260603
  \urlprefix\url{https://link.aps.org/doi/10.1103/PhysRevLett.115.260603}

\bibitem{Scandi2019}
Scandi M and Perarnau-Llobet M 2019 {\em Quantum\/} {\bf 3} 197
  \urlprefix\url{https://doi.org/10.22331/q-2019-10-24-197}

\bibitem{Brandner2020}
Brandner K and Saito K 2020 {\em Phys. Rev. Lett.\/} {\bf 124} 040602
  \urlprefix\url{https://link.aps.org/doi/10.1103/PhysRevLett.124.040602}

\bibitem{Frim2022}
Frim A~G and DeWeese M~R 2022 {\em Phys. Rev. Lett.\/} {\bf 128} 230601
  \urlprefix\url{https://link.aps.org/doi/10.1103/PhysRevLett.128.230601}

\bibitem{Miller2020}
Miller H~J~D and Mehboudi M 2020 {\em Phys. Rev. Lett.\/} {\bf 125} 260602
  \urlprefix\url{https://link.aps.org/doi/10.1103/PhysRevLett.125.260602}

\bibitem{Erdman2022}
Erdman P~A, Rolandi A, Abiuso P, Perarnau-Llobet M and Noé F 2022
  Pareto-optimal cycles for power, efficiency and fluctuations of quantum heat
  engines using reinforcement learning
  \urlprefix\url{https://arxiv.org/abs/2207.13104}

\bibitem{Lu2022}
Lu J, Wang Z, Peng J, Wang C, Jiang J~H and Ren J 2022 {\em Phys. Rev. B\/}
  {\bf 105} 115428
  \urlprefix\url{https://link.aps.org/doi/10.1103/PhysRevB.105.115428}

\bibitem{Bhandari2020}
Bhandari B, Alonso P~T, Taddei F, von Oppen F, Fazio R and Arrachea L 2020 {\em
  Phys. Rev. B\/} {\bf 102} 155407
  \urlprefix\url{https://link.aps.org/doi/10.1103/PhysRevB.102.155407}

\bibitem{Potanina2021}
Potanina E, Flindt C, Moskalets M and Brandner K 2021 {\em Phys. Rev. X\/} {\bf
  11} 021013
  \urlprefix\url{https://link.aps.org/doi/10.1103/PhysRevX.11.021013}

\bibitem{Eglinton2022}
Eglinton J and Brandner K 2022 {\em Phys. Rev. E\/} {\bf 105} L052102
  \urlprefix\url{https://link.aps.org/doi/10.1103/PhysRevE.105.L052102}

\bibitem{Izumida2021}
Izumida Y 2021 {\em Phys. Rev. E\/} {\bf 103} L050101
  \urlprefix\url{https://link.aps.org/doi/10.1103/PhysRevE.103.L050101}

\bibitem{Izumida2022}
Izumida Y 2022 {\em Phys. Rev. Res.\/} {\bf 4} 023217
  \urlprefix\url{https://link.aps.org/doi/10.1103/PhysRevResearch.4.023217}

\bibitem{Alonso2022}
Terr\'en~Alonso P, Abiuso P, Perarnau-Llobet M and Arrachea L 2022 {\em Phys.
  Rev. X Quantum\/} {\bf 3} 010326
  \urlprefix\url{https://link.aps.org/doi/10.1103/PRXQuantum.3.010326}

\bibitem{Raz2016}
Raz O, Suba\ifmmode \mbox{\c{s}}\else \c{s}\fi{}\ifmmode \imath \else~\i \fi{}
  Y and Pugatch R 2016 {\em Phys. Rev. Lett.\/} {\bf 116} 160601
  \urlprefix\url{https://link.aps.org/doi/10.1103/PhysRevLett.116.160601}

\bibitem{Bhandari2022}
Bhandari B and Jordan A~N 2022 {\em Phys. Rev. Res.\/} {\bf 4} 033103
  \urlprefix\url{https://link.aps.org/doi/10.1103/PhysRevResearch.4.033103}

\bibitem{Frim2022b}
Frim A~G and DeWeese M~R 2022 {\em Phys. Rev. E\/} {\bf 105} L052103
  \urlprefix\url{https://link.aps.org/doi/10.1103/PhysRevE.105.L052103}

\bibitem{Hino2021}
Hino Y and Hayakawa H 2021 {\em Phys. Rev. Res.\/} {\bf 3} 013187
  \urlprefix\url{https://link.aps.org/doi/10.1103/PhysRevResearch.3.013187}

\bibitem{Hayakawa2021}
Hayakawa H, Paasonen V~M~M and Yoshii R 2021 Geometrical quantum chemical
  engine \urlprefix\url{https://arxiv.org/abs/2112.12370}

\bibitem{Chen2022}
Chen J~F 2022 {\em Phys. Rev. E\/} {\bf 106} 054108
  \urlprefix\url{https://link.aps.org/doi/10.1103/PhysRevE.106.054108}

\bibitem{Pietzonka2018}
Pietzonka P and Seifert U 2018 {\em Phys. Rev. Lett.\/} {\bf 120} 190602
  \urlprefix\url{https://link.aps.org/doi/10.1103/PhysRevLett.120.190602}

\bibitem{Chen2017}
Chen Y~Y, Watanabe G, Yu Y~C, Guan X~W and del Campo A 2019 {\em npj Quant.
  inf.\/} {\bf 5} \urlprefix\url{https://doi.org/10.1038/s41534-019-0204-5}

\bibitem{Li2018}
Li J, Fogarty T, Campbell S, Chen X and Busch T 2018 {\em New J. Phys.\/} {\bf
  20} 015005 \urlprefix\url{https://doi.org/10.1088/1367-2630/aa9cd8}

\bibitem{Boubakour2022}
Boubakour M, Fogarty T and Busch T 2022 Interaction enhanced quantum heat
  engine \urlprefix\url{https://arxiv.org/abs/2211.03394}

\bibitem{Niedenzu2018}
Niedenzu W and Kurizki G 2018 {\em New J. Phys.\/} {\bf 20} 113038
  \urlprefix\url{https://doi.org/10.1088/1367-2630/aaed55}

\bibitem{Kloc2019}
Kloc M, Cejnar P and Schaller G 2019 {\em Phys. Rev. E\/} {\bf 100} 042126
  \urlprefix\url{https://link.aps.org/doi/10.1103/PhysRevE.100.042126}

\bibitem{Yadin2022}
Yadin B, Morris B and Brandner K 2022 Thermodynamics of permutation-invariant
  quantum many-body systems: A group-theoretical framework
  \urlprefix\url{https://arxiv.org/abs/2206.12639}

\bibitem{Kolisnyk2022}
Kolisnyk D and Schaller G 2022 Performance boost of a collective qutrit
  refrigerator \urlprefix\url{https://arxiv.org/abs/2210.07844}

\bibitem{Carollo2020}
Carollo F, Brandner K and Lesanovsky I 2020 {\em Phys. Rev. Lett.\/} {\bf 125}
  240602
  \urlprefix\url{https://link.aps.org/doi/10.1103/PhysRevLett.125.240602}

\bibitem{jaseem2022}
Jaseem N, Vinjanampathy S and Mukherjee V 2022 Quadratic enhancement in the
  reliability of collective quantum engines
  \urlprefix\url{https://arxiv.org/abs/2208.04250}

\bibitem{Ma2017}
Ma Y~H, Su S~H and Sun C~P 2017 {\em Phys. Rev. E\/} {\bf 96} 022143
  \urlprefix\url{https://link.aps.org/doi/10.1103/PhysRevE.96.022143}

\bibitem{Fogarty_2021}
Fogarty T and Busch T 2020 {\em Quantum Sci. Technol.\/} {\bf 6} 015003
  \urlprefix\url{https://dx.doi.org/10.1088/2058-9565/abbc63}

\bibitem{wen2022}
Wen X, Fan R and Vishwanath A 2022 Floquet's refrigerator: Conformal cooling in
  driven quantum critical systems
  \urlprefix\url{https://arxiv.org/abs/2211.00040}

\bibitem{Revathy2022}
BS R, Mukherjee V and Divakaran U 2022 {\em Entropy\/} {\bf 24}
  \urlprefix\url{https://www.mdpi.com/1099-4300/24/10/1458}

\bibitem{Myres2020}
Myers N~M and Deffner S 2020 {\em Phys. Rev. E\/} {\bf 101} 012110
  \urlprefix\url{https://link.aps.org/doi/10.1103/PhysRevE.101.012110}

\bibitem{Myres2021}
Myers N~M, McCready J and Deffner S 2021 {\em Symmetry\/} {\bf 13}
  \urlprefix\url{https://www.mdpi.com/2073-8994/13/6/978}

\bibitem{Myres2022}
Myres N~M, Peña F~J, Negrete O, Vargas P, Chiara G~D and Deffner S 2022 {\em
  New J. Phys.\/} {\bf 24} 025001
  \urlprefix\url{https://doi.org/10.1088/1367-2630/ac47cc}

\bibitem{Koch2022}
Koch J, Menon K, Cuestas E, Barbosa S, Lutz E, Fogarty T, Busch T and Widera A
  2022 Making statistics work: a quantum engine in the bec-bcs crossover
  \urlprefix\url{https://arxiv.org/abs/2209.14202}

\bibitem{marzolino2022}
Marzolino U 2022 Quantum thermochemical engines
  \urlprefix\url{https://arxiv.org/abs/2208.04132}

\bibitem{Skelt2019}
Skelt A~H, Zawadzki K and D'Amico I 2019 {\em J. Phys. A: Math. Theor.\/} {\bf
  52}

\bibitem{Carollo2020b}
Carollo F, Gambetta F~M, Brandner K, Garrahan J~P and Lesanovsky I 2020 {\em
  Phys. Rev. Lett.\/} {\bf 124} 170602
  \urlprefix\url{https://link.aps.org/doi/10.1103/PhysRevLett.124.170602}

\bibitem{Mukherjee2021}
Mukherjee V and Divakaran U 2021 {\em J. Phys.: Condens. Matter\/} {\bf 33}
  454001 \urlprefix\url{https://doi.org/10.1088/1361-648x/ac1b60}

\bibitem{Mayo2022}
Mayo F and Roncaglia A~J 2022 {\em Phys. Rev. A\/} {\bf 105} 062203
  \urlprefix\url{https://link.aps.org/doi/10.1103/PhysRevA.105.062203}

\bibitem{Li2022}
Li J, Sherman E~Y and Ruschhaupt A 2022 {\em Phys. Rev. A\/} {\bf 106} L030201
  \urlprefix\url{https://link.aps.org/doi/10.1103/PhysRevA.106.L030201}

\bibitem{solfanelli2022}
Solfanelli A, Giachetti G, Campisi M, Ruffo S and Defenu N 2022 Quantum heat
  engine with long-range advantages
  \urlprefix\url{https://arxiv.org/abs/2208.09492}

\bibitem{Ketterle1996}
Ketterle W and van Druten N~J 1996 {\em Phys. Rev. A\/} {\bf 54} 656--660
  \urlprefix\url{https://doi.org/10.1103/PhysRevA.54.656}

\bibitem{pitaevskii2003}
Pitaevskii L, P{\'\i}tajevsk{\'\i}j L, S~Stringari L, Stringari S and Stringari
  S 2003 {\em Bose-Einstein Condensation\/} International Series of Monographs
  on Physics (Clarendon Press)

\bibitem{Alicki1979}
Alicki R 1979 {\em J. Phys. A: Math. Gen\/} {\bf 12} L103--L107
  \urlprefix\url{https://doi.org/10.1088/0305-4470/12/5/007}

\bibitem{Spohn1978b}
Spohn H 1978 {\em J. Math. Phys.\/} {\bf 19} 1227--1230
  \urlprefix\url{https://doi.org/10.1063/1.523789}

\bibitem{Spohn1978}
Spohn H and Lebowitz J~L 1978 {\em Irreversible Thermodynamics for Quantum
  Systems Weakly Coupled to Thermal Reservoirs\/} (John Wiley $\&$ Sons, Ltd)
  pp 109--142
  \urlprefix\url{https://onlinelibrary.wiley.com/doi/abs/10.1002/9780470142578.ch2}

\bibitem{Brandner2016}
Brandner K and Seifert U 2016 {\em Phys. Rev. E\/} {\bf 93} 062134
  \urlprefix\url{https://doi.org/10.1103/PhysRevE.93.062134}

\bibitem{Schmitt2014}
Schmitt J, Damm T, Dung D, Vewinger F, Klaers J and Weitz M 2014 {\em Phys.
  Rev. Lett.\/} {\bf 112} 030401
  \urlprefix\url{https://link.aps.org/doi/10.1103/PhysRevLett.112.030401}

\bibitem{Crisanti2019}
Crisanti A, Sarracino A and Zannetti M 2019 {\em Phys. Rev. Res.\/} {\bf 1}
  023022
  \urlprefix\url{https://link.aps.org/doi/10.1103/PhysRevResearch.1.023022}

\end{thebibliography}

\end{document}